\documentclass[12pt,preprint]{emulateapj}

\usepackage{color}

\slugcomment{Accepted: May 3, 2011}

\shorttitle{Dust formation in SNe Ia}
\shortauthors{Nozawa et al.}

\begin{document}

\title{Formation of Dust in the Ejecta of Type I\lowercase{a} 
Supernovae}

\author{Takaya Nozawa,\altaffilmark{1} Keiichi Maeda,\altaffilmark{1}
 Takashi Kozasa,\altaffilmark{2} Masaomi Tanaka,\altaffilmark{1} \\
 Ken'ichi Nomoto,\altaffilmark{1} and Hideyuki Umeda\altaffilmark{3}}

\altaffiltext{1}{Institute for the Physics and Mathematics of the
Universe, University of Tokyo, Kashiwa, Chiba 277-8583, Japan; 
takaya.nozawa@ipmu.jp}
\altaffiltext{2}{Department of Cosmosciences, Graduate School 
of Science, Hokkaido University, Sapporo 060-0810, Japan}
\altaffiltext{3}{Department of Astronomy, School of Science, University
of Tokyo, Bunkyo-ku, Tokyo 113-0033, Japan}

\begin{abstract}
We investigate the formation of dust grains in the ejecta of Type Ia 
supernovae (SNe Ia), adopting the carbon-deflagration W7 model.
In the calculations of dust formation, we apply the nucleation and 
grain growth theory and consider the two cases with and without 
formation of CO and SiO molecules.
The results of the calculations show that for the sticking probability 
of $\alpha_j =$ 1, C, silicate, Si, and FeS grains can condense at 
early times of $\sim$100--300 days after the explosion, whereas Fe and 
SiC grains cannot form substantially.
Due to the low gas density in SNe Ia with no H-envelope, the average 
radii of the newly formed grains are generally below 0.01 $\mu$m, being
much smaller than those in Type II-P SNe.
This supports our previous conclusion that the radius of dust formed 
in the ejecta is smaller in SNe with less massive envelopes. 
The total dust mass ranges from $3 \times 10^{-4}$ $M_{\odot}$ to 0.2 
$M_{\odot}$ for $\alpha_j =$ 0.1--1, depending on whether or not CO and 
SiO molecules are formed.
We also estimate the optical depths and thermal emission by the newly 
formed dust and compare to the relevant observations of SNe Ia.
We find that the formation of C grains in SNe Ia is suppressed to be 
consistent with the observational constraints.
This implies that energetic photons and electrons heavily depress the 
formation efficiency of C grains or that the outermost C-O layer of SNe 
Ia is almost fully burned.
Finally, we perform the calculations of dust destruction in the 
SN remnants and find that dust grains formed in the ejecta of 
SNe Ia are almost completely destroyed in the shocked gas before being 
injected into the interstellar medium. 
This indicates that SNe Ia are unlikely to be the major sources of 
interstellar dust.

\end{abstract}

\keywords{dust, extinction --- galaxies: abundances --- infrared: stars
--- ISM: supernova remnants --- supernovae: general --- white dwarfs}

\section{Introduction}

Type Ia supernovae (SNe Ia) are considered to be thermonuclear explosions 
of carbon-oxygen white dwarfs (WDs).
In the single degenerate scenario, the progenitor of an exploding WD 
grows up to a mass close to the Chandrasekhar limit through mass 
transfer from the binary companion and finally undergoes carbon 
ignition near the center (e.g., Nomoto 1982).
However, it has been a matter of controversy whether the nuclear burning 
front propagates as a subsonic deflagration wave (Nomoto et al.\ 1976, 
1984b) or as a supersonic detonation wave growing from the deflagration 
(Khokhlov 1991a, 1991b).
Although there are some differences in the resulting elemental 
distributions, both the propagation modes can synthesize a significant 
amount of iron-peak elements as well as intermediate-mass elements such 
as Si, S, and Ca (H\"{o}flich et al.\ 1995; Iwamoto et al.\ 1999).
Thus, SNe Ia are major sources of heavy elements and play a critical
role in the chemical enrichment history of the universe (e.g., Nomoto 
et al.\ 1984a; Timmes et al.\ 1995; Kobayashi \& Nomoto 2009).

It has been also supposed that SNe Ia can be possible producers of 
dust grains, especially Fe grains (Tielens 1998) and would have 
important consequences for the evolution and inventory of interstellar 
dust (Dwek 1998). 
This presumption is mainly based on their metal-rich compositions of the 
cooling gas in the ejecta similar to those in core-collapse supernovae 
(CCSNe) for which several pieces of observational evidence for dust 
formation have been reported (see Kozasa et al.\ 2009 for review).
In addition, the formation of dust in SNe Ia has been suggested from 
the analysis of presolar grains extracted from meteorites;
Clayton et al.\ (1997) pointed out that the isotopic signatures of type 
X silicon carbide (SiC) particles can be explained if SiC grains condense 
out of the explosive helium burning layer in SNe Ia.
However, the composition, size, and amount of dust grains formed in the 
ejecta of SNe Ia and injected into the interstellar medium (ISM) have not 
been fully explored to date.

It should be mentioned that systematic studies of SN Ia rates have 
proposed the presence of a prompt component of SNe Ia exploding in a 
timescale as short as 0.1 Gyr after the stellar birth (Mannucci et 
al.\ 2005, 2006; Della Valle et al.\ 2005; Scannapieco \& Bildsten 2005).
The increasing testimonies of such short-lived SNe Ia have been 
thereafter given by many theoretical and observational works 
(Totani et al.\ 2008; Aubourg et al.\ 2008; Hachisu et al.\ 2008a, 
2008b; Matteucci et al.\ 2009; Brandt et al.\ 2010, Maoz et al.\ 2010).
Although in the single degenerate scenario SNe Ia are not expected to 
generate in metal-poor environments as low as [Fe/H] $< -1.1$ 
(Kobayashi et al.\ 1998), prompt SNe Ia might have occur in the early 
universe at redshifts higher than $z=4$, where most of observed quasar 
systems have already reached solar- and super-solar metallicities 
(Juarez et al.\ 2009).
Hence, if prompt SNe Ia could actually produce dust grains, they would 
have made a large contribution to dust budgets not only in our Galaxy 
but also in high-redshift dusty galaxies (e.g., Valiante et al.\ 2009).

On the other hand, no observation of normal SNe Ia has reported 
convincing evidence for the ongoing formation of dust in the ejecta
as indicated by an increase of infrared (IR) continuum, a rapid decline 
of optical luminosity, and blueshifts of atomic line emissions;
Gerardy et al.\ (2007) observed two SNe Ia, SN 2003hv and SN 2005df, 
at $t \sim$ 100--400 days since the explosion with the {\it Spitzer 
Space Telescope}.
However, the observed mid-IR spectral energy distributions (SEDs) did 
not show any evolution, allowing the authors to conclude that the mid-IR 
fluxes are dominated by the strong atomic line emission rather than 
thermal emission from newly formed dust.
Maeda et al.\ (2009) discovered an accelerated fading of the optical 
light curve for the peculiar SN Ia 2006gz, but it remains unclear whether 
this is responsible for dust formation.
Furthermore, no IR emission from dust formed in the ejecta has been 
seen in the supernova remnant (SNR) of Tycho (Douvion et al.\ 2001) 
identified as a standard SN Ia (Krause et al.\ 2008), although recently 
Ishihara et al.\ (2010) suggested the possible detection of thermal 
emission from the ejecta-origin dust, based on the IR observations with 
the {\it AKARI}.

The fact that the existence of newly formed dust has not been distinctly
confirmed so far implies that formation of dust in SNe Ia is likely to 
be inefficient or impossible because of some physical conditions in the 
ejecta different from CCSNe;
SNe Ia have much higher expansion velocities of the ejecta and produce 
much more radioactive element $^{56}$Ni than typical CCSNe.
A higher expansion velocity leads to a lower gas density in the ejecta, 
which prevents dust grains from growing to radii larger than 0.01 
$\mu$m or even makes the condensation of dust unsuccessful (Nozawa et 
al.\ 2008, 2010).
In addition, energetic photons and electrons generated from the decay 
of radioactive elements can inhibit the formation of dust grains 
(Nozawa et al.\ 2008) and also can disturb the formation of SiO and CO 
molecules that affects the dust formation processes 
(H\"{o}flich et al.\ 1995).
Thus, it is essential to examine how the formation process of dust in 
the ejecta depends on the type of SNe, by taking account of the time 
evolution of gas temperature and density, the formation of CO and SiO 
molecules, and the impacts of high-energy photons and electrons.

In this paper, we investigate the formation of dust in the expanding 
ejecta of SNe Ia, with the aim of revealing the roles of SNe Ia in the 
dust enrichment in the universe. 
In Section 2, we describe the models of SNe Ia and the method of 
calculation of dust formation.
In the calculations, we adopt as a model of SNe Ia the 
carbon-deflagration W7 model (Nomoto et al.\ 1984b; Thielemann et al.\ 
1986), and apply the nucleation and grain growth theory to estimate the 
composition, size, and mass of dust that can condense in the ejecta.
The results of the calculations are presented in Section 3 for the two 
cases with and without the formation of CO and SiO molecules.
In Section 4, we calculate the optical depths and thermal emissions by 
the newly formed dust and apply them to the observational constraints.
Then we discuss the formation process of dust in SNe Ia and the 
elemental composition in the outermost layer of SNe Ia.
In Section 5, we calculate the destruction of the newly formed grains
by the reverse shock and examine whether they can survive to be 
injected from SNe Ia into the ISM.
The summary is devoted in Section 6.

\section{Calculation of Dust Formation in the Ejecta of SN\lowercase{e} 
I\lowercase{a}}

\subsection{Model of SNe Ia}

The composition, size, and number density of dust formed in SNe depends 
on the elemental composition in the ejecta as well as the time 
evolution of the density and temperature of the gas (Kozasa et al.\ 1989, 
1991; Nozawa et al.\ 2003).
In the calculations of dust formation in SNe Ia, we adopt the 
nucleosynthesis and hydrodynamic results of the one-dimension 
carbon-deflagration W7 model (Nomoto et al.\ 1984b; Thielemann et al.\ 
1986).
The W7 model can account for many of the observed characteristics of 
normal SNe Ia, so it has been extensively used as one of the most 
standard SN Ia models.
In the W7 model, the ejecta mass is $M_{\rm eje} = 1.38$ $M_\odot$, 
and the kinetic energy of the explosion is $E_{\rm kin}/10^{51}$erg
$=E_{51}=1.3$.
The mass of synthesized $^{56}$Ni is $M$($^{56}$Ni) $= 0.6$ $M_\odot$, 
which is one order of magnitude larger than those ($M$($^{56}$Ni) 
$\simeq 0.06$ $M_\odot$) produced in typical CCSNe.

%%%%%%%%%%%%%%%%%%%%%%%%%%%%%%%%%%
\begin{figure}
%\epsscale{0.6}
\epsscale{1.0}
\plotone{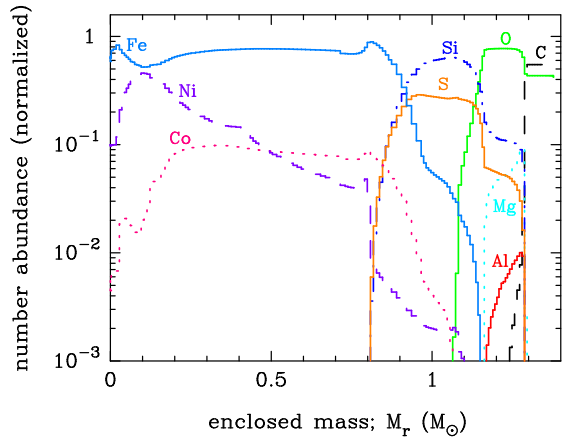}
\caption{
 Relative number abundances of elements taking part in dust formation in 
 the ejecta of the SN Ia (W7 model) at 300 days after the explosion as 
 a function of enclosed mass.
 The decay of radioactive elements is taken into account, and the abundance
 of each element is summed up over the isotopes.
 In the calculations of dust formation, the original onion-like 
 composition is assumed to be preserved with no mixing of elements.
\label{fig1}}
\end{figure}
%%%%%%%%%%%%%%%%%%%%%%%%%%%%%%%%%%

Figure 1 shows the number abundances of elements relevant to dust 
formation in the ejecta of the W7 model at 300 days after the explosion
as a function of enclosed mass $M_r$.
In Type II-P SNe (SNe II-P) with massive H-envelopes, dust grains can 
only form in the metal-rich gas inside the He core (Kozasa et al.\ 1989).
However, SNe Ia do not have H-rich envelopes, and dust formation would be 
possible in the entire region of the ejecta.
We divide the ejecta of the W7 model into four layers, according to the 
elemental composition of interest to dust formation:
the Fe--Ni layer ($M_r =$ 0--0.91 $M_\odot$), 
the Si--S layer ($M_r =$ 0.91--1.14 $M_\odot$), 
the O-rich layer ($M_r =$ 1.14--1.28 $M_\odot$), 
and the C--O layer ($M_r =$ 1.28--1.38 $M_\odot$).

In the deflagration W7 model, 0.1 $M_\odot$ of C and O remains unburned 
in the outermost layer.
In W7, the initial C+O white dwarf is {\sl assumed} to be composed of 
$X(^{12} {\rm C}) = 0.475$, $X(^{16} {\rm O}) = 0.50$, and 
$X(^{22} {\rm Ne}) = 0.025$ (where $X$ denotes the mass fraction)
with a number ratio of C/O $= 1.27$.
Therefore, we can expect the formation of C grains in the outermost 
C--O layer, although the C/O ratio in the outermost layer is determined 
by many cycles of recurrent He-shell flashes, thus being highly uncertain.
On the other hand, in the delayed detonation model, pre-existing C atoms 
can be completely burned, except for those in the very outer layers with 
the expansion velocities higher than $\sim$20,000 km s$^{-1}$ (Iwamoto 
et al.\ 1999), which prohibits the condensation of large amounts of C 
grains.
The formation of C grains and the elemental composition in the outer 
layer of SNe Ia will be discussed in Section 4.4.

In the calculations, we consider the onion-like elemental composition 
as shown in Figure 1, assuming that any microscopic and macroscopic 
mixing of elements do not occur before the condensation of dust grains.
This assumption is supported by a variety of studies;
the spectrum synthesis calculations of a time-series of spectra obtained 
during $t =$5--377 days for SN 2002bo, SN 2004eo, and SN 2003du 
demonstrated that the abundance distributions in the ejecta of these 
SNe Ia are fully stratified with some degree of macroscopic mixing 
(Stehle et al.\ 2005; Mazzali et al.\ 2008; Tanaka et al.\ 2011).
Based on the analysis of the emission profiles in the mid-IR spectrum at 
$t \simeq 135$ days, Gerardy et al.\ (2007) have also shown that the 
ejecta structure in SN Ia 2005df is chemically stratified.
Furthermore, X-ray observations of the Tycho SNRs indicate that the 
layered composition has been retained in the ejecta even at $\sim$450 yr 
after the explosion (Badenes et al.\ 2006; Hayato et al.\ 2010 see also 
Kosenko et al.\ 2010 for Type Ia SNR 0519--69.0).

The time evolution of the gas density in the freely expanding ejecta is 
calculated as $\rho \propto t^{-3}$, based on the density structure of 
the W7 model.
The time evolution of the gas temperature is evaluated by solving 
the radiative transfer equation simultaneously with the energy equation 
of the radiation plus gas under the assumption of the local thermodynamic 
equilibrium (LTE, Iwamoto et al.\ 2000). 
The energy deposition through the decay of radioactive isotopes is 
calculated by giving a constant $\gamma$-ray absorption opacity that 
can reproduce the behavior of the bolometric light curves observed for 
standard SNe Ia.
The kinetic energies of positrons are assumed to be deposited 
instantaneously in situ.
The energy loss by positron escape becomes more important at later times 
when the density in the ejecta is low enough.
However, at the epochs relevant to this study ($t \la 300$ days), this 
effect can be negligible (Milne et al.\ 2001).

Figures 2(a) and 2(b) show the structures of the gas temperature and 
density in the ejecta of the W7 model, respectively, at 100 days 
(thick solid lines) and 300 days (thick dashed lines) after 
the explosion.
Figure 2(c) displays the velocity distribution of the W7 model.
For comparison, we also show those in the He core of the SN II-P model 
with an $M_{\rm eje} = 17$ $M_\odot$, an $E_{51}=1$, and a massive 
H-envelope of $M_{\rm Henv} = 13.2$ $M_\odot$ (thin lines; Umeda \& 
Nomoto 2002).
As can be seen from Figure 2(c), the expansion velocity of the 
metal-rich ejecta is much higher in the SN Ia than in the SN II-P, 
despite the almost same explosion energy of both the SNe;
in the SN Ia all of the explosion energy is deposited into the low-mass 
ejecta with no H-envelope, whereas in the SN II-P most of explosion 
energy is used to blow off the massive H-envelope.
As a result, the gas density in the SN Ia is more than three orders of 
magnitude lower than that in the SN II-P.
In addition, the gas temperature in the SN Ia decreases more quickly 
than that in the SN II-P and drops down to $\sim$2000 K at 100 days.
This is because the lower gas density causes absorption of $\gamma$-ray 
in the ejecta to be less efficient and because the thermal energy 
produced at the explosion is quickly lost by the adiabatic expansion 
unlike SNe II-P.

%%%%%%%%%%%%%%%%%%%%%%%%%%%%%%%%%%
\begin{figure}
%\epsscale{0.6}
\epsscale{1.0}
\plotone{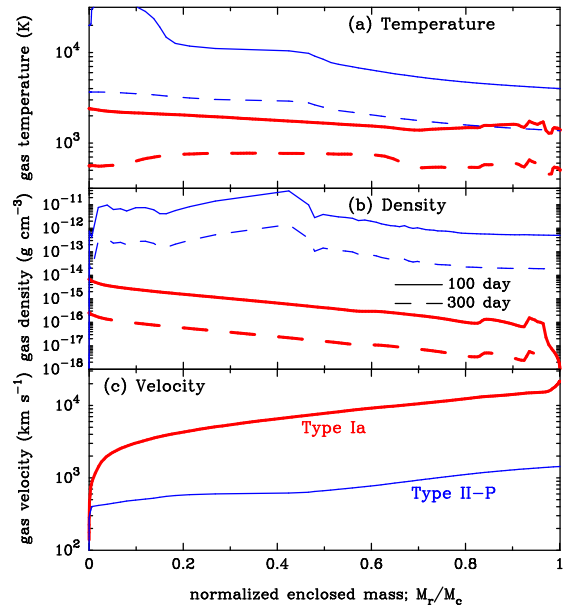}
\caption{
 (a) Temperature and (b) density distributions of the gas in the ejecta 
 of the SN Ia (W7 model) at day 100 (thick solid lines) and day 300 
 (thick dashed lines) after the explosion, and (c) the velocity structure 
 for the W7 model (thick lines).
 The enclosed mass is normalized to the ejecta mass of the SN Ia
 ($M_{\rm c} = M_{\rm eje} = 1.38$ $M_\odot$). 
 For comparison, shown are those for the SN II-P model (thin lines) with 
 an $M_{\rm pr} = 20$ $M_\odot$ and an $E_{51} = 1$ (Umeda \& Nomoto 2002). 
 The mass coordinate for the SN II-P is normalized to the mass of the 
 metal-rich ejecta ($M_{\rm c} = M_{\rm He core} - M_{\rm mass cut} = 
 3.34$ $M_\odot$) within which dust formation is realized (Nozawa et al.\ 
2003).
\label{fig2}}
\end{figure}
%%%%%%%%%%%%%%%%%%%%%%%%%%%%%%%%%%

\subsection{Nucleation Rate}

One of the main aims of this paper is to clarify how the size and mass 
of dust formed in the ejecta depend on the type of SNe.
In order to achieve this goal, we apply a non-steady state nucleation 
and grain growth theory, which has been utilized for investigating the 
dust formation in various types of CCSNe in our previous studies 
(Nozawa et al.\ 2003, 2008, 2010).
The theory enables us to estimate the size distribution and mass of newly 
formed dust, given the elemental composition and the time evolution of 
density and temperature of the gas.
Here we present the formula for the nucleation rate, which will be 
useful for discussion in the later section.
For comprehensive description of dust formation calculations, we refer
the readers to Nozawa et al.\ (2003).
We also note that there is another approach to evaluate the amount of 
dust formed in the SN ejecta, which follows the abundances of dust
precursor molecules by solving a complex chemical reaction network
(Cherchneff \& Dwek 2009, 2010).

%%%%%%%%%%%%%%%%%%%%%%%%%%%%%%%%%%
\begin{deluxetable*}{clclcc}
\tablewidth{0pt}
\tablecaption{Chemical reactions for silicate grains considered in 
this paper}
\tablehead{ 
& \colhead{dust species} & 
\colhead{key species} & \colhead{chemical reaction} & 
\colhead{$A/10^4$K} & \colhead{$B$} 
}
\startdata
\multicolumn{6}{c}{model A} \\
\tableline
(1) & MgSiO$_3$$_{\rm (s)}$     & Mg$_{\rm (g)}$/SiO$_{\rm (g)}$  &
Mg$_{\rm (g)}$ + SiO$_{\rm (g)}$ + 2O$_{\rm (g)}$ 
$\rightarrow$ MgSiO$_3$$_{\rm (s)}$  & 25.0129  & 72.0015 \\
(2) & Mg$_2$SiO$_4$$_{\rm (s)}$ & Mg$_{\rm (g)}$  &
2Mg$_{\rm (g)}$ + SiO$_{\rm (g)}$ + 3O$_{\rm (g)}$ 
$\rightarrow$ Mg$_2$SiO$_4$$_{\rm (s)}$  & 18.6200 & 52.4336  \\
    & & SiO$_{\rm (g)}$  &
2Mg$_{\rm (g)}$ + SiO$_{\rm (g)}$ + 3O$_{\rm (g)}$ 
$\rightarrow$ Mg$_2$SiO$_4$$_{\rm (s)}$  & 37.2400 & 104.8672 \\
(3) & SiO$_2$$_{\rm (s)}$       & SiO$_{\rm (g)}$  &
SiO$_{\rm (g)}$ + O$_{\rm (g)}$ 
$\rightarrow$ SiO$_2$$_{\rm (s)}$  & 12.6028 & 38.1507 \\ \hline
\tableline
 \multicolumn{6}{c}{model B} \\
\tableline
(4) & MgSiO$_3$$_{\rm (s)}$     & Mg$_{\rm (g)}$/Si$_{\rm (g)}$  &
Mg$_{\rm (g)}$ + Si$_{\rm (g)}$ + 3O$_{\rm (g)}$ 
$\rightarrow$ MgSiO$_3$$_{\rm (s)}$  & 34.7214  & 87.7178 \\
(5) & Mg$_2$SiO$_4$$_{\rm (s)}$ & Mg$_{\rm (g)}$  &
2Mg$_{\rm (g)}$ + Si$_{\rm (g)}$ + 4O$_{\rm (g)}$ 
$\rightarrow$ Mg$_2$SiO$_4$$_{\rm (s)}$  & 23.4742 & 60.2918 \\
    &  & Si$_{\rm (g)}$  &
2Mg$_{\rm (g)}$ + Si$_{\rm (g)}$ + 4O$_{\rm (g)}$ 
$\rightarrow$ Mg$_2$SiO$_4$$_{\rm (s)}$  & 46.9484 & 120.5836 \\
(6) & SiO$_2$$_{\rm (s)}$       & Si$_{\rm (g)}$  &
Si$_{\rm (g)}$ + 2O$_{\rm (g)}$ 
$\rightarrow$ SiO$_2$$_{\rm (s)}$  & 22.3113 & 53.8670 \\
\enddata
\tablecomments{
The key species is defined as the gas species whose collisional 
frequency is the least among reactants.
The subscripts (s) and (g) denote solid condensate and gas species, 
respectively.
The Gibbs free energy $\Delta G^0_j$ for the formation of a condensate 
from reactants per molecule of the key species is approximated 
by $\Delta G^0_j / k T_{d,j} = - A/T_{d,j} + B$, where the numerical 
values $A$ and $B$ are derived by least-squares fittings of the 
thermodynamics data (Chase et al.\ 1985) in the range of temperature of
interests.}
\end{deluxetable*}
%%%%%%%%%%%%%%%%%%%%%%%%%%%%%%%%%%

The nucleation rate, taking account of chemical reactions at 
condensation, is evaluated by introducing the concept of the key 
species defined as the least abundant gaseous species among the 
reactants.
By assuming that the key species controls the kinetics of nucleation and 
grain growth (Kozasa \& Hasegawa 1987; Hasegawa \& Kozasa 1988) and taking 
into account the difference between the temperatures of gas and 
condensation nuclei (Kozasa et al.\ 1996), the steady-state nucleation 
rate $J^s_j$ for a given grain species $j$ is written as\footnote
{There is a typographical error in Eq (3) for nucleation rate in 
Nozawa et al.\ (2003); $c_{1,j}$ should be replaced with $c_{1,j}^2$
as is seen in Eq (1).}
\begin{eqnarray}
J^s_j = \alpha_j \Omega_j \left( \frac{2 \sigma_j}{\pi m_{1,j}}
\right)^{\frac{1}{2}} \left( \frac{T}{T_{d,j}} \right)^{\frac{1}{2}}
c_{1,j}^2 \exp \left[- \frac{4 \mu^3_j}{27 (\ln S_j)^2} \right],
\end{eqnarray}
where $\alpha_j$ is the probability with which the key species stick onto
the surface of grains, $\mu_j = 4 \pi a_{0,j}^2 \sigma_j / k T_{d,j}$ is 
the energy barrier for nucleation, $k$ is the Boltzmann constant, 
$\sigma_j$ is the surface energy, $T$ is the gas temperature, $T_{d,j}$ 
is the temperature of the condensation nuclei, and
$\Omega_j$ and $a_{0,j}= (3 \Omega_j / 4 \pi)^{1/3}$ 
are, respectively, the volume and hypothetical radius of the condensate 
per molecule of the key species. 
The concentration and mass of the key species are represented as 
$c_{1,j}$ and $m_{1,j}$, respectively.
Note that a factor $\Pi_j$, which depends on the partial pressures of 
reactant and product gas species except for the key species (Chigai et 
al.\ 1999), is omitted from the equation since we set $\Pi_j = 1$ in the 
calculations (see Nozawa et al.\ 2003).
The supersaturation ratio $S_j$ is calculated by 
\begin{eqnarray}
\ln S_j = - \frac{\Delta G^0_j}{k T_{d,j}} + \sum_i \nu_{i} \ln P_{i}
+ \frac{1}{2} \ln \left( \frac{T}{T_{d,j}} \right),
\end{eqnarray}
where $\Delta G^0_j$ is the Gibbs free energy for condensation of the 
$j$th dust species from the reactants per molecule of the key species, 
and $P_{i}$ are the partial gas pressures of the reactants and products.
The stoichiometric coefficients $\nu_{i}$, which are normalized to the 
key species, are positive for a reactant and negative for a product.

In the following calculations, we assume $T_{d,j} = T$, to be consistent 
with a treatment in our earlier studies;
dust formation in the case where the temperature of the condensation 
nuclei is not equal to the gas temperature is examined in Section 4.2.
We take $\alpha_j =$ 1 and 0.1 as the sticking probability, assuming 
that it is the same for all grain species; 
for $\alpha_j \le$ 0.01, dust formation does not occur substantially 
for the SN model considered here. 
The Gibbs free energy for formation of a grain species is approximated 
as $\Delta G^0_j/k T_{d,j} = - A/T_{d,j} + B$ through the least-square 
fitting of the thermodynamics data (Chase et al.\ 1985).
The formation of all possible condensates is calculated simultaneously.
The grain species considered and the data needed for the dust formation 
calculations are given in Table 2 in Nozawa et al.\ (2003).

\subsection{Formation of CO and SiO Molecules}

Since the first detection in SN 1987A (Spyromilio et al.\ 1988; Aitken 
et al.\ 1988), formation of CO and SiO molecules has been observed in 
several CCSNe (Spyromilio \& Leibundgut 1996; Spyromilio et al.\ 2001; 
Gerardy et al.\ 2000, 2002; Kotak et al.\ 2006, 2009; Hunter et al.\ 
2009). 
Formation of these molecules prior to dust formation can affect the 
composition, size, and mass of dust grains that condense in the ejecta
(e.g., Nozawa et al.\ 2003);
carbon (oxygen) atoms bound in CO molecules are not available for
formation of C-bearing (O-bearing) grains, and SiO molecules lock up 
silicon atoms available to form Si and SiC. 
On the other hand, SiO molecules, whose rapid depletion was observed 
concurrently with the onset of dust formation in SN 1987A (Roche et 
al.\ 1991) and SN 2004et (Kotak et al.\ 2009), are considered to be 
precursors of silicate grains such as MgSiO$_3$, Mg$_2$SiO$_4$, and 
SiO$_2$ (Kozasa et al.\ 1989) as represented by the reactions (1)--(3) 
in Table 1.

In the astronomical environments hospitable to dust formation such as 
in mass-loss winds of evolved late-type stars, it is usually assumed 
that formation of CO and SiO molecules is complete.
However, in the ejecta of SNe, these molecules are destroyed by impacts
with energetic electrons and/or charge transfer reactions with the 
ionized inert gas.
For SN 1987A, the masses of CO and SiO molecules were estimated to be 
$\la 5 \times 10^{-3}$ $M_\odot$ (Liu et al.\ 1992; Liu \& Dalgarno 
1995; Gearhart et al.\ 1999) and $\la 10^{-3}$ $M_\odot$ (Liu \& 
Dalgarno 1994, 1996), respectively.
These masses correspond to formation efficiencies less than a few 
percents, where the formation efficiency is defined as the mass 
fraction of carbon and silicon locked in CO and SiO molecules.
The disruption of CO molecules may enable carbon grains to condense out 
of the oxygen-rich gas in the SN ejecta (Clayton et al.\ 1999, 2001; 
Deneault et al.\ 2006; Todini \& Ferrara 2001; Bianchi \& Schneider 
2007).

In the ejecta of SNe Ia that produce much more $^{56}$Ni than typical 
CCSNe, destruction of CO and SiO molecules can be more effective as a 
result of more vigorous radioactivity (Liu 1997).
H\"{o}flich et al.\ (1995) argued that CO and SiO molecules can form 
more or less in subluminous SNe Ia with $M$($^{56}$Ni) $\la 0.3$ 
$M_\odot$ but cannot form in normal SNe Ia with $M$($^{56}$Ni) 
$\simeq 0.6$ $M_\odot$.
Indeed, neither the fundamental nor overtone emissions of CO molecule 
have been detected in ordinary SNe Ia, although Taubenberger et al.\ 
(2010) has reported probable detection of CO emission at $\sim$85 day 
after the peak brightness in superluminous SN Ia 2009dc.
We also note that possible formation of SiO molecule is supposed for 
SN Ia 2005df, but its formation efficiency should be extremely low 
($\la$ 0.01 \%, Gerardy et al.\ 2007).

In the present calculations, we consider two extreme cases for 
formation of CO and SiO molecules.
One is the complete formation of CO and SiO molecules, in which case the 
lesser of carbon (silicon) and oxygen is locked up in CO (SiO) molecules
and SiO molecules act as precursors of silicate grains.
This is the assumption made by a series of our previous works.
Although CO and SiO molecules can be the dominant coolants in SNe 
(Liu \& Dalgarno 1995), we do not take into account the effects of 
cooling by these molecules on the thermal structure of the 
ejecta.\footnote{
The cooling by CO and SiO molecules causes rapid decreases in the gas 
temperatures in the C-O layer and the O-rich layer.
This would lead to the earlier condensation of silicate and C grains 
and their larger average radii than those given in this paper.
However, the temperature decrease by molecular cooling cannot
cause the mass of these dust grains to increase appreciably in the case 
of $\alpha_j = 1$, where most of the metals available for dust 
formation are incorporated into dust grains (see Section 3.1). 
We also note that if the ejecta were optically thick for CO and SiO 
emissions, the cooling could not operate effectively (Gearhart et al.\ 
1999).}
The other case is no formation of CO and SiO molecules, where carbon, 
silicon, and oxygen take part in dust formation as free atoms.
It should be mentioned here that even if SiO molecules cannot form in 
the ejecta, silicate grains may condense via the chemical reaction paths 
different from the reactions (1)--(3) given in Table 1.
Here we calculate the formation of silicates by considering the 
reactions (4)--(6) in Table 1.

\section{Results of Dust Formation Calculations}

In this section, we present the results obtained from the models of 
calculations outlined in Section 2.
In what follows, we refer to the cases with and without formation of CO 
and SiO molecules as the model A and model B, respectively, with the 
values of sticking probabilities $\alpha_j$ attached in the model names.
For example, the A1 represents the model taking complete CO and SiO 
formation with $\alpha_j = 1$.
For each of models considered in this paper, the mass of each dust 
species formed in the ejecta are summarized in Table 2.
In the followings, we focus on the results in the most optimistic 
cases of $\alpha_j = 1$ for dust formation (i.e., the models A1 and B1), 
to make a direct comparison with our previous studies, where 
$\alpha_j = 1$ was assumed.

%%%%%%%%%%%%%%%%%%%%%%%%%%%%%%%%%%
\begin{deluxetable*}{lcccc}
\tablewidth{0pt}
\tablecaption{Mass of Each Dust Species Formed in the Ejecta of the 
SN\lowercase{e} I\lowercase{a} (W7 model)}
\tablehead{ 
\colhead{dust species} & 
\colhead{A1} & \colhead{A0.1} & 
\colhead{B1} & \colhead{B0.1}
}
\startdata
C             & $5.66 \times 10^{-3}$ & $2.84 \times 10^{-4}$
              & $3.73 \times 10^{-2}$ & $2.40 \times 10^{-2}$   \\
MgO           & $3.17 \times 10^{-6}$ & $1.85 \times 10^{-9}$
              & $9.26 \times 10^{-8}$ & $1.93 \times 10^{-9}$   \\
MgSiO$_3$     & $7.59 \times 10^{-3}$ & $1.31 \times 10^{-6}$
              & $1.95 \times 10^{-2}$ & $1.11 \times 10^{-5}$   \\
Mg$_2$SiO$_4$ & $7.01 \times 10^{-3}$ & $1.50 \times 10^{-6}$
              & $6.08 \times 10^{-3}$ & $6.49 \times 10^{-6}$   \\
SiO$_2$       & $1.47 \times 10^{-2}$ & $9.94 \times 10^{-6}$
              & $4.91 \times 10^{-2}$ & $2.21 \times 10^{-3}$   \\
Al$_2$O$_3$   & $8.18 \times 10^{-7}$ & $7.48 \times 10^{-10}$
              & $8.53 \times 10^{-6}$ & $7.71 \times 10^{-10}$  \\
FeS           & $1.78 \times 10^{-2}$ & $1.53 \times 10^{-5}$
              & $1.78 \times 10^{-2}$ & $1.53 \times 10^{-5}$   \\
Si            & $6.30 \times 10^{-2}$ & $3.15 \times 10^{-5}$
              & $6.40 \times 10^{-2}$ & $3.21 \times 10^{-5}$   \\
Fe            & $9.52 \times 10^{-5}$ & $1.09 \times 10^{-8}$
              & $9.52 \times 10^{-5}$ & $1.09 \times 10^{-8}$   \\
Ni            & $1.48 \times 10^{-6}$ & $2.22 \times 10^{-10}$
              & $1.48 \times 10^{-6}$ & $2.22 \times 10^{-10}$  \\ \hline
Total         & $1.16 \times 10^{-1}$ & $3.44 \times 10^{-4}$
              & $1.94 \times 10^{-1}$ & $2.63 \times 10^{-2}$   
\enddata
\tablecomments{
The dust mass is given in units of $M_\odot$.
The models A1 and A0.1 are the cases where formation of CO and SiO 
molecules is assumed to be complete, with the sticking probabilities of 
$\alpha_{j}=$ 1 and 0.1, respectively.
In the model B1 with $\alpha_{j}=$ 1 and the model B0.1 with 
$\alpha_{j}=$ 0.1, it is assumed that any molecule never forms in the 
ejecta.}
\end{deluxetable*}
%%%%%%%%%%%%%%%%%%%%%%%%%%%%%%%%%%

\subsection{Cases with Molecular Formation}

Figure 3(a) shows the condensation time $t_c$ of dust grains formed in 
the ejecta of SNe Ia for the model A1, where $t_c$ is defined as the 
time when the nucleation rate reaches the maximum (see Nozawa et al.\ 
2003).
In the outermost C--O layer, C grains are formed at 80--110 days from 
the remaining carbon atoms that are not locked in CO molecules.
In the O-rich layer, Al$_2$O$_3$ grains first condense around 100 days, 
and silicate (Mg$_2$SiO$_4$, MgSiO$_3$, SiO$_2$) grains condense at 
120--150 days via the reactions involving SiO molecules.
Then MgO grains form at 150--160 days in the O-rich layer, and FeS and 
Si grains form at 160--250 days in the Si--S layer.
Despite the presence of the extended Fe--Ni layer, Fe and Ni grains 
are not produced at $M_r =$ 0.15--0.8 $M_\odot$; 
the gas density in this region becomes too low for them to nucleate 
before the gas cools down to their condensation temperatures 
($\la$ 800 K).
Fe and Ni grains can condense at 250--300 days only in the innermost 
region ($M_r =$ 0--0.15 $M_\odot$) with the highest gas density 
(see Figure 2), though their masses are very small (see below).

%%%%%%%%%%%%%%%%%%%%%%%%%%%%%%%%%%
\begin{figure}
%\epsscale{0.6}
\epsscale{1.0}
\plotone{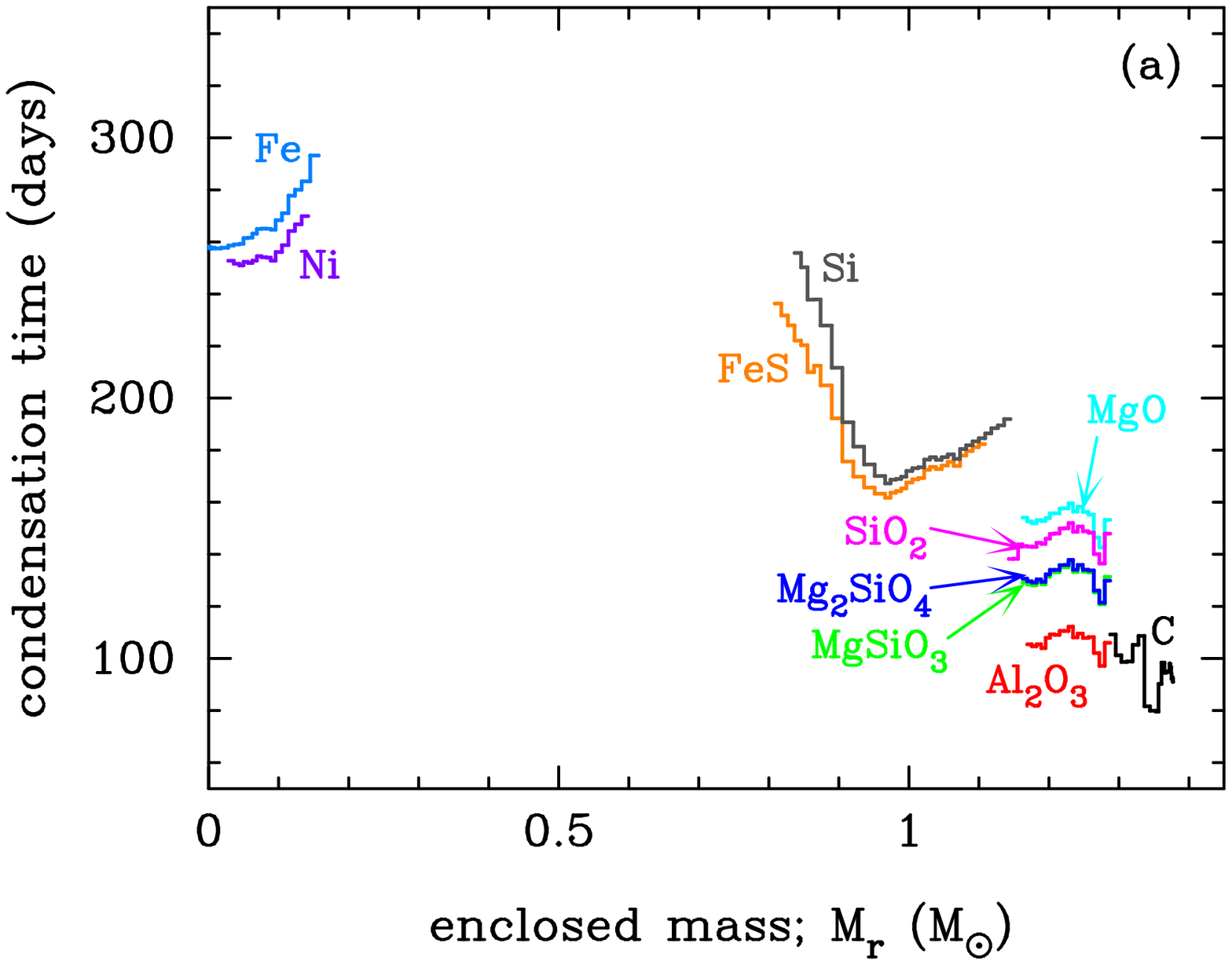}
\plotone{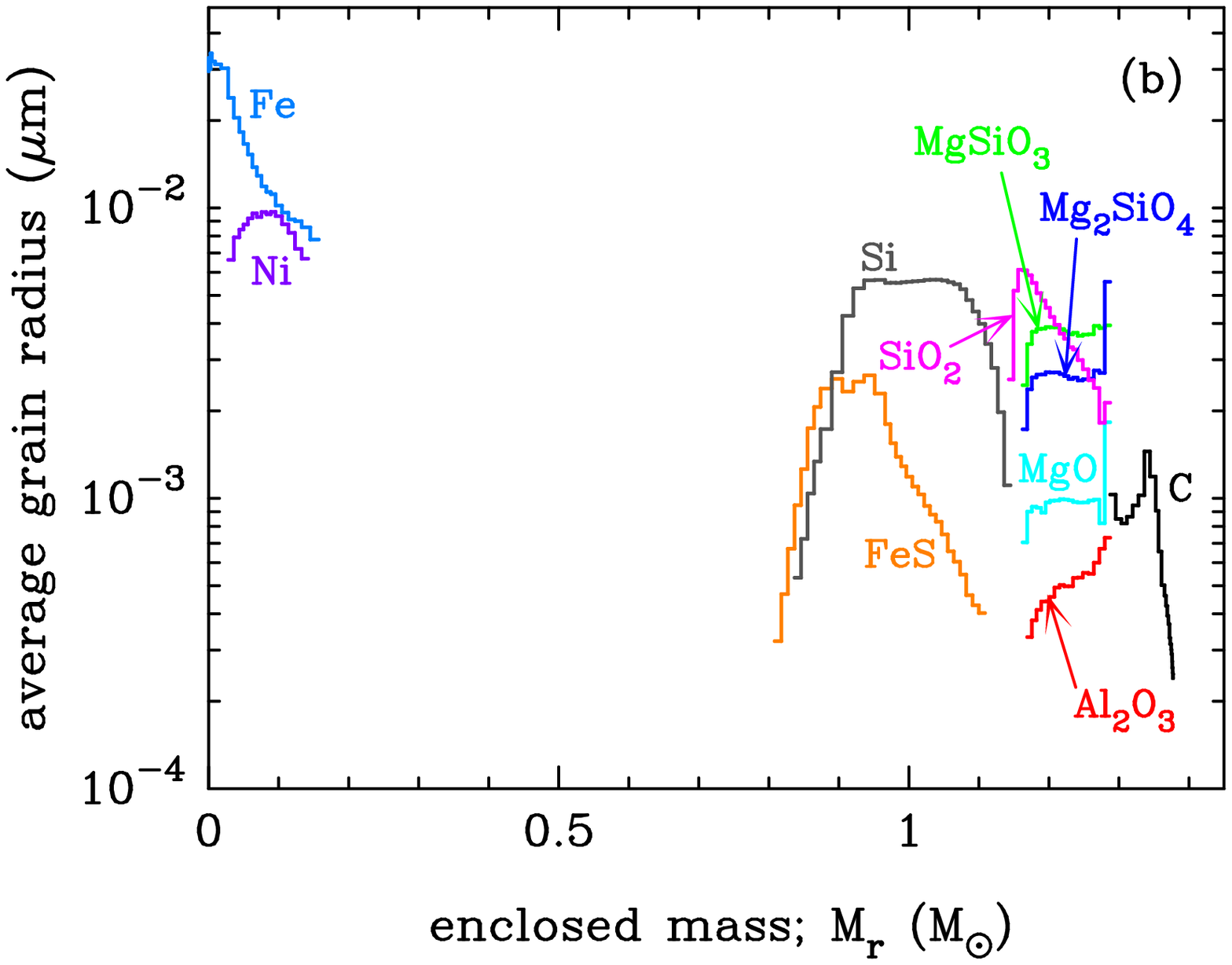}
\caption{
 (a) Condensation times and (b) average radii of dust grains 
 formed in the ejecta of the W7 model as a function of enclosed mass for 
 the model A1 taking complete formation of CO and SiO molecules with
 $\alpha_j = 1$.
\label{fig3}}
\end{figure}
%%%%%%%%%%%%%%%%%%%%%%%%%%%%%%%%%%

The above results indicate that a variety of dust grains can form in SNe 
Ia according to the elemental abundances in the different layers, given 
that the stratified elemental composition is retained with no mixing.
The resulting composition of newly formed grains and the order of their 
condensation are almost the same as those in any types of CCSNe 
(Nozawa et al.\ 2003, 2008, 2010).
This means that the condensation sequence of dust does not depend on the
type of SNe, reflecting the fact that the elemental composition of the 
metal-rich gas in the ejecta is not greatly different among different 
types of SNe.
However, the condensation times of dust in SNe Ia are much earlier 
($t_c \simeq$ 100--300 days) than those ($t_c \ga$ 300 days) in SNe II-P 
(Nozawa et al.\ 2003).
Such early condensation of dust is caused by the rapid decrease of 
the gas temperature resulting from the low gas density in the ejecta 
(see Figure 2).

The low gas density in SNe Ia also leads to an important consequence 
for the size of dust formed in the ejecta.
Figure 3(b) exhibits the average radii $a_{\rm ave}$ of newly formed 
grains as a function of enclosed mass.
We can see that the average radii of all dust species except Fe and Ni 
grains are under 0.01 $\mu$m, being considerably smaller than those 
($a_{\rm ave} \ga 0.01$ $\mu$m) in SNe II-P (Nozawa et al.\ 2003).
As demonstrated by Nozawa et al.\ (2010), the condensation of dust 
grains in less dense gas demands a larger supercooling, and the
resulting formation and growth of a huge number of condensation nuclei 
makes the average radius of grains small due to the conservation of mass 
of materials available for dust formation.
On the other hand, in the gas with an even much lower density, formation 
of seed nuclei itself is depressed significantly.
In this case, the final radius is determined by the competition between
the collision timescale of the key species onto the grain surfaces and 
the expansion timescale of the gas.
This is the case for Fe and Ni grains, which can acquire the average 
radii of $a_{\rm ave} \sim 0.01$ $\mu$m

It should be emphasized that the masses of Fe and Ni grains are quite 
low ($\la 10^{-4}$ $M_\odot$) although their radii are relatively 
large. 
This indicates that SNe Ia cannot be major sources of Fe and Ni grains, 
contrary to the presupposition made in some studies.
In the model A1, the total mass of newly formed dust is 0.116 
$M_\odot$ and the main dust species in mass are Si (0.063 $M_\odot$), 
FeS (0.018 $M_\odot$), and SiO$_2$ grains (0.015 $M_\odot$).
The mass of silicate (Mg$_2$SiO$_4$, MgSiO$_3$, SiO$_2$) grains is 
0.03 $M_\odot$ in total and occupies 25 \% of the total dust mass. 
C grains are produced with $5.7 \times 10^{-3}$ $M_\odot$. 
In the model A0.1 with $\alpha_j = 0.1$, the total dust mass is 
$3.4 \times 10^{-4}$ $M_\odot$ and is dominated by C grains (see 
Table 2).
The average grain radii are a factor of 5--10 smaller in the model A0.1 
than in the model A1.
However, the condensation times of dust grains for the model A0.1 are 
not significantly different from those for the model A1.

\subsection{Cases with No Molecular Formation}

Figures 4(a) and 4(b) show the condensation times and the average 
radii of newly formed grains in the model B1, where no molecular 
formation and $\alpha_j = 1$ are assumed.
Since dust formation in the inner Fe--Ni layer ($M_r <$ 0.8 $M_\odot$) 
is not affected by formation of CO and SiO molecules, only shown the 
results in the region of $M_r =$ 0.8--1.38 $M_\odot$ are in Figure 4. 
In the model B1 with no CO formation, nucleation and growth of C grains 
in the outermost C--O layer can advance in more carbon-rich gas.
As a result, the average radii and the mass of C grains increase by a 
factor of 2--3 and 6.6, respectively, compared with those in the model A1.
Furthermore, because of no formation of SiO molecules, more free silicon 
atoms are available for formation of Si grains at $M_r =$ 1.10--1.15 
$M_\odot$, leading to a little higher mass of Si grains than in the model 
A1.

%%%%%%%%%%%%%%%%%%%%%%%%%%%%%%%%%%
\begin{figure}
%\epsscale{0.6}
\epsscale{1.0}
\plotone{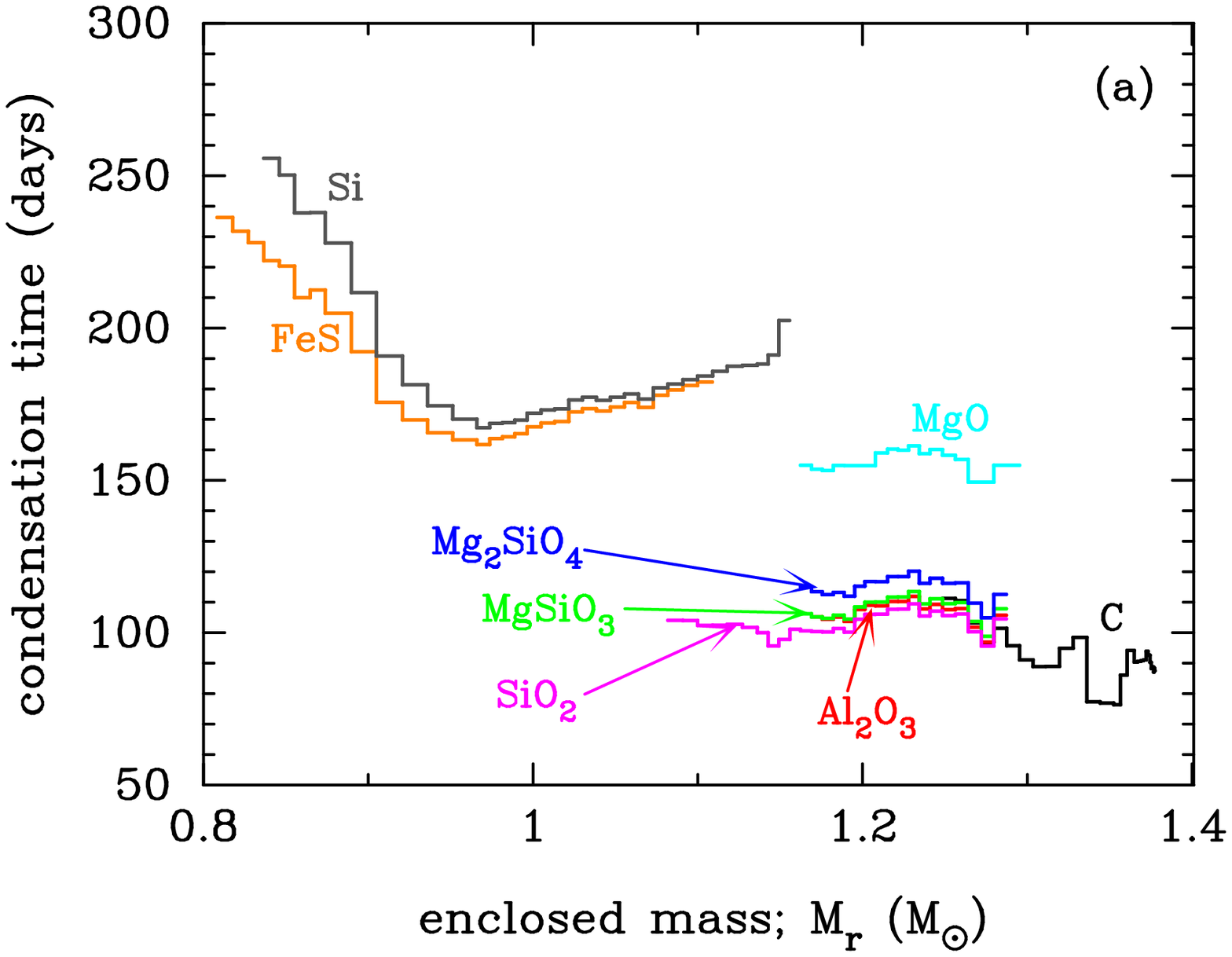}
\plotone{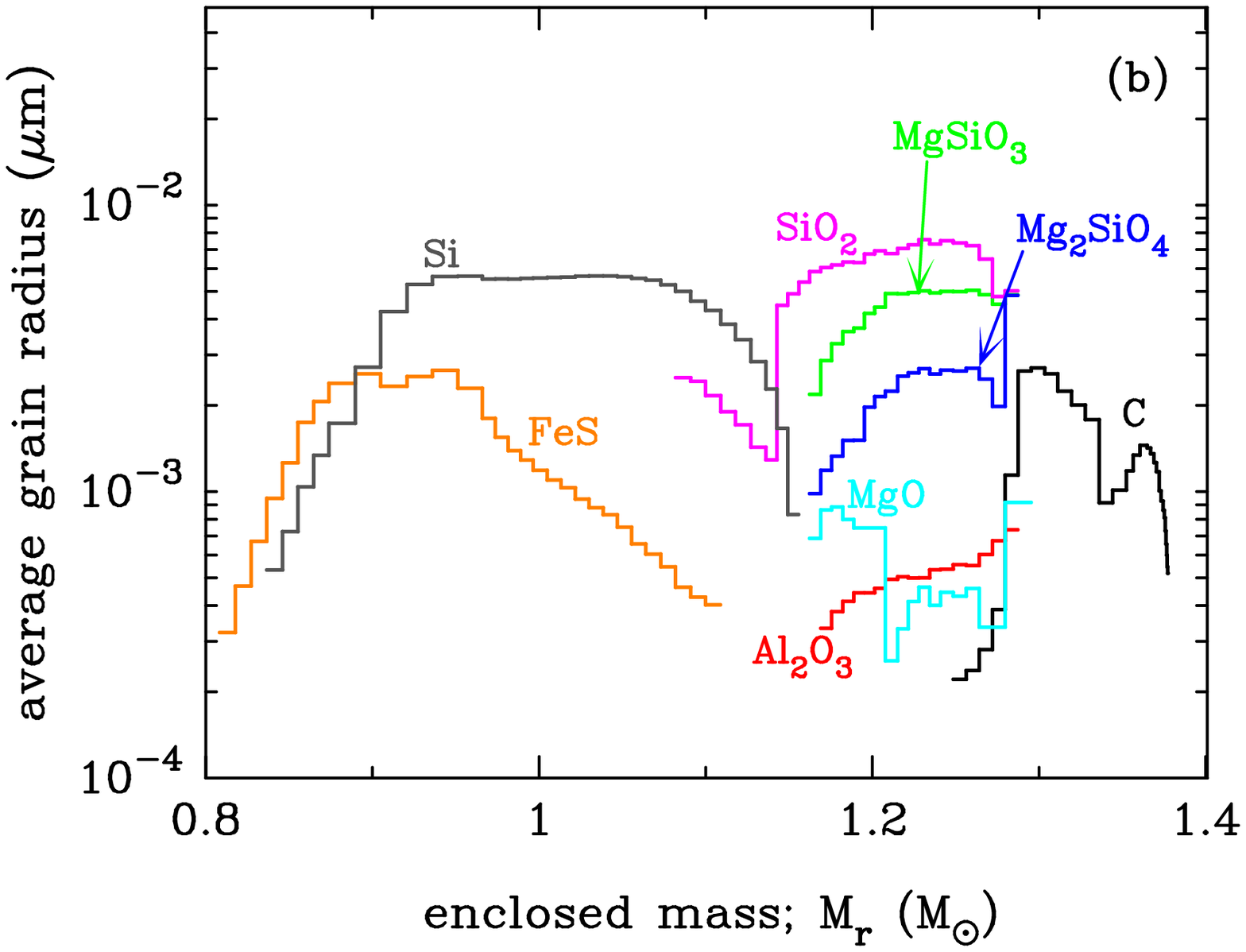}
\caption{
 (a) Condensation times and (b) average radii of dust grains formed 
 at $M_r =$ 0.8--1.38 $M_\odot$ in the W7 model for the model B1 taking
 no formation of CO and SiO molecules with $\alpha_j = 1$.
\label{fig4}}
\end{figure}
%%%%%%%%%%%%%%%%%%%%%%%%%%%%%%%%%%

The difference of the results between the models A1 and B1 also appears 
in formation of silicate grains.
In the model B1, Mg$_2$SiO$_4$, MgSiO$_3$, and SiO$_2$ grains can 
condense a few tens days earlier than those in the model A1.
Moreover, the condensation time of SiO$_2$ grains is earlier than those
of Mg$_2$SiO$_4$ and MgSiO$_3$, in contrast to the case in the model A1.
These differences comes from the difference in the Gibbs free energies 
for their condensation;
the reactions (4)--(6) in Table 1 provide the lower Gibbs free energies 
than the reactions (1)--(3).
Since the lower Gibbs free energy allows the supersaturation ratio to 
rise above unity more rapidly (see Eq (2)) for a given temperature 
evolution of cooling gas, the earlier formation of silicate, especially 
SiO$_2$ grains, is realized in the model B.
The average radii and the masses are enhanced for SiO$_2$ and MgSiO$_3$, 
as a result of their formation in the denser gas (at earlier times).
However, both the average radius and the mass are slightly reduced for 
Mg$_2$SiO$_4$, compared with those in the model A1;
the earlier condensation of SiO$_2$ and MgSiO$_3$ significantly reduces 
Si atoms available for the later formation of Mg$_2$SiO$_4$.
The total mass of silicate is 0.075 $M_\odot$, which is 2.5 times 
higher than those in the model A1.

Our calculations demonstrate that SiC grains never condense in SNe Ia, 
regardless of formation of CO and SiO molecules.
The main reason for this is that in the ejecta of W7 model there is no 
layer where carbon and silicon coexist abundantly.
However, even if both the elements are abundant in some layer, as 
appeared in CCSNe, theoretical calculations (Nozawa et al.\ 2003; 
Cherchneff \& Dwek 2009, 2010) do not predict the condensation of SiC 
grains with radii more than 0.1 $\mu$m as discovered in presolar grains.
Thus, it remains unexplained whether presolar SiC grains originated 
from SNe Ia (Clayton et al.\ 1997) or CCSNe (Amari et al.\ 1992; 
Nittler et al.\ 1996, see also Deneault et al.\ 2003; Deneault 2009).

In summary, formation of CO and SiO molecules, as well as the sticking 
probability, has substantial impacts on the mass of dust grains formed 
in SNe Ia;
the total dust mass for the model B1 (B0.1) without molecular formation 
is 0.194 $M_\odot$ (0.026 $M_\odot$) and is much higher than 0.116 
$M_\odot$ ($3 \times 10^{-4}$ $M_\odot$) for the model A1 (A0.1) with 
molecular formation.
In particular, the difference in dust mass is more than two orders of 
magnitude between the models of A0.1 and B0.1 for which the total dust 
masses are dominated by C grains.
In the model B0.1 without CO formation, the number density of C atoms 
available for dust formation is about five times higher than in the 
model A0.1.
Thus, even if $\alpha_j = 0.1$, condensation of C grains can still 
proceed efficiently, and a considerable amount of C grains can form in 
the model B0.1.

On the other hand, in all of the models considered here, the 
condensation times of dust are significantly early ($<$ 300 days), 
and the average grain radii are small ($\la$ 0.01 $\mu$m).
These conclusions are not influenced by CO and SiO formation and 
arise from the low density of the gas in the ejecta of SNe Ia with no 
H-envelope.
Our preceding studies demonstrated that a tiny size of dust grains 
with $a_{\rm ave} \la 0.01$ $\mu$m condense in envelope-stripped Type Ib 
SN 2006jc (Nozawa et al.\ 2008) and Type IIb SNe (Nozawa et al.\ 2010), 
while relatively large grains with $a_{\rm ave} \ga 0.01$ $\mu$m form 
in envelope-retaining SNe II-P (Nozawa et al.\ 2003).
Thus, we conclude that the radius of dust formed in the ejecta depends 
on the type of SNe, and that smaller dust grains are produced as the 
envelope masses of SNe become lower.

\section{Discussion}

As shown in the last section, dust grains of 
$\simeq 3 \times 10^{-4}$--0.2 $M_\odot$ can form in the ejecta of SNe 
Ia for the sticking probabilities of $\alpha_j =$ 0.1--1.
The formation of such an amount of dust grains may show some 
observational signatures such as a fading of optical luminosity and a 
rising of infrared luminosity.
However, these events suggestive of dust formation have not been 
confirmed by any observation of ordinary SNe Ia.
In this section, we calculate the opacity and thermal emission by 
the newly formed dust and apply the results to the observational 
constraints.
Then we discuss the composition and mass of dust formed in the ejecta 
as well as the elemental composition of the outermost layer in SNe Ia.

\subsection{Optical Depth by Newly Formed Dust}

We begin with estimating the optical depths by dust formed in the 
ejecta.
The optical depth $\tau_\lambda$ as measured from the outermost radius 
$R$ of the ejecta is calculated as follows,
\begin{eqnarray}
\tau_\lambda(r,t)
&=& \sum_j \tau_{\lambda,j}(r,t) \nonumber \\
&=& \sum_j \int^r_R dr' \int^{a_{{\rm max},j}}_{a_{{\rm min},j}} 
m_j(a) \kappa^{\rm ext}_{\lambda, j}(a) f_j(a, r', t) da,
\end{eqnarray}
where $f_j(a, r, t)$ is the size distribution function of dust species 
$j$ per volume at a position $r$, $m_j(a) = 4 \pi a^3 \rho_j/ 3$ is the 
mass of a grain with a radius $a$ and a bulk density $\rho_j$, and
$\kappa^{\rm ext}_{\lambda, j}(a)$ is the mass extinction coefficient.
The maximum and minimum radii of dust ($a_{{\rm max},j}$ 
and $a_{{\rm min},j}$, respectively), as well as $f_j(a, r, t)$, are 
obtained from the dust formation calculations in Section 3.
Here we assume that dust grains formed at a give position distribute 
homogeneously within the corresponding mesh.
The bulk density $\rho_j$ of each dust species is taken from Table 2 
in Nozawa et al.\ (2006), and the optical constants for calculating 
$\kappa^{\rm ext}_{\lambda, j}$ are taken from the references summarized 
in Section 4.1 in Nozawa et al.\ (2010).\footnote
{The bulk density and the source of optical constants for Ni grains
have not been given in Nozawa et al.\ (2006, 2010).
We adopt $\rho_{\rm Ni} = 8.87$ g cm$^{-3}$ as a bulk density of Ni 
grains, which is evaluated by using the hypothetical radius 
$a_{0,j} = 1.377$ \AA~of the condensate (Nozawa et al.\ 2003).
For the optical constants, we assume that the optical properties of 
Ni grains are the same as those of Fe grains.
Since the mass of Ni grains formed in the ejecta is negligible, this 
assumption has no impact on the results in the rest of this paper.} 
Note that the optical depths scale as $\tau_\lambda \propto t^{-2}$ 
for the free expansion of the ejecta.

The results of calculations show that the total optical depths at V band 
($\lambda = 0.55$ $\mu$m) are $\tau_{0.55} =$ 163 (279) at 300 days for 
the model A1 (B1); 
C, Si, FeS, and Fe grains are the major donors of the opacity with 
$\tau_{0.55,j} =$ 22 (137), 78 (79), 14 (14), and 49 (49), respectively.
At 300 days, silicate grains only have $\tau_{0.55,{\rm sil}} =$ 0.01 
(0.02) because of their low extinction efficiencies at $\lambda = 0.55$ 
$\mu$m, while they provide large optical depths of $\tau_{10,j} =$ 5.4 
(11.7) at $\lambda =$ 10 $\mu$m.
Note that even for the model A0.1 (B0.1) with $\alpha_j = 0.1$, the 
V-band optical depths by C grains reach $\tau_{0.55} =$ 2.5 (93) at 
200 days, which are large enough to affect the optical light curve of 
the SN.
The calculation by Sugerman et al.\ (2006) suggested that the clumpy 
structure of the ejecta can make the opacity by dust smaller by about 
one order of magnitude than the homogeneous distribution with the same 
dust mass (see also Ercolano et al.\ 2007).
However, even if we consider the effect of dust clump, the V-band
opacities calculated here are still too high to be consistent with the 
observations.
This indicates that the mass of dust formed in SNe Ia must be smaller
than our calculations predict.

SNe Ia produce approximately ten times more $^{56}$Ni than normal 
CCSNe and thus may have stronger radiation fields in the ejecta.
Moreover, dust formation in SNe Ia is realized within the first one 
year, during which energetic photons and electrons are likely to be 
abundant inside the ejecta.
In such an environment, small clusters that are composed of several to 
a few tens atoms are heated up by the impacts with UV-optical photons 
and fast electrons, and their temperature could be higher than the 
temperature of the gas, depending on their optical properties.
The heating of small clusters can lead to detachment of atoms from the 
clusters and thus can inhibit the formation of dust.
In the next subsection, we evaluate the temperature of condensate 
nuclei, and examine how the formation process of dust can be affected 
by the effect of non-local thermal equilibrium between the nucleated 
dust and the gas.

\subsection{Dust Formation in non-Local Thermal Equilibrium Condition}

In the calculations of dust formation in Section 3, we assumed that
the temperature of condensation nuclei $T_{d,j}$ is the same as the 
temperature of the gas $T$.
However, $T_{d,j}$ is not generally equal to $T$, and the resulting 
temperature difference can influence the condensation process of dust 
(Kozasa et al.\ 1996).
In particular, if $T < T_{d,j}$ in the cooling gas, the supersaturation 
ratio ($\ln S_j$) increases more slowly than the case of the local 
thermal equilibrium with $T = T_{d, j}$ (see Eq (2)), which retards the 
formation of dust.
Furthermore, the effects of the enhanced detachment of monomers from 
a heated small cluster can be seen in Equation (1) via the factors 
$(T / T_{d,j})^{1/2}$ and $\mu_j \propto T_{d,j}^{-1}$.

The temperature of a condensate is determined by the balance between 
the heating and the cooling.
In the ejecta of SNe, the possible heating processes of small clusters 
are absorption of UV-optical radiation and collisions with energetic 
electrons, while the cooling of dust takes place via thermal emission 
and collisions with cooler gaseous atoms.
However, the energy transferred through collisions with the gas is found 
to be much lower than that through radiative processes.
Thus, we neglect the heating and cooling by collisions with electrons 
and gaseous atoms.
Also, for simplicity, we do not consider the effect of stochastic 
heating in evaluating the temperature of small clusters;
in the SN ejecta with a relatively strong radiation field, the 
temperature of a stochastically heated grain distributes symmetrically 
around the equilibrium temperature (Nozawa et al.\ 2008), 
which therefore does not affect significantly the IR emission spectra
calculated in the next subsection as well.

By balancing the heating by photon absorption and the cooling by
thermal radiation of dust, the equilibrium temperature of dust 
$T_{d,j}(r,t)$ at a given time $t$ and at a given position $r$ 
is calculated from the equation 
\begin{eqnarray}
F(r,t) K_{F,j}(a,T_*) = 4 \sigma_{\rm B} T_{d,j}^4(r,t) K_{P,j}(a,T_{d,j}),
\end{eqnarray}
where $F(r,t)$ is the flux, $\sigma_{\rm B}$ is the Stefan-Boltzmann 
constant, and $K_{F,j}(a,T)$ and $K_{P,j}(a,T)$ are, respectively,  
the flux-mean and the Planck-mean of absorption coefficient 
$Q_{\lambda,j}(a)$; here we assume that the optical properties of small 
clusters are the same as those of the small dust grains.
The flux of the radiation field $F(r,t)$ in the ejecta has been obtained 
from the radiative transfer calculation in deriving the time evolution 
of the gas temperature.
In the SN model considered here, the flux at day 400 ranges from 
$F = 1.7 \times 10^5$ erg s$^{-1}$ cm$^{-2}$ to 
$F = 8.2 \times 10^5$ erg s$^{-1}$ cm$^{-2}$ 
over most parts of the ejecta.
Note that the spectral energy distribution of the radiation field in the 
ejecta at late nebula phases have not been known very well both 
observationally and theoretically.
In the calculations, we assume that the radiation field has a blackbody 
spectrum with a temperature $T_* = T_{\rm BB} = 5000$ K, not depending 
on position and time.

%%%%%%%%%%%%%%%%%%%%%%%%%%%%%%%%%%
\begin{figure}
%\epsscale{0.7}
\epsscale{1.1}
\plotone{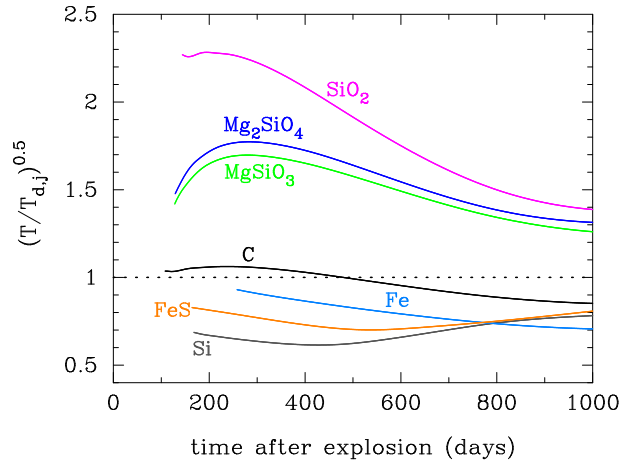}
\caption{
 Time evolutions of the square root of the ratio of the gas temperature
 to the dust temperature ($T / T_{d,j})^{1/2}$ for the main dust
 species:
 Fe grains at $M_r =$ 0.001 $M_\odot$, 
 Si and FeS grains at $M_r =$ 0.98 $M_\odot$, 
 MgSiO$_3$, Mg$_2$SiO$_4$, and SiO$_2$ grains at $M_r =$ 1.18 $M_\odot$, 
 and C grains at $M_r =$ 1.33 $M_\odot$.
\label{fig5}}
\end{figure}
%%%%%%%%%%%%%%%%%%%%%%%%%%%%%%%%%%

In Figure 5, we present the time evolutions of the square root of the 
ratio between the gas temperature and the dust temperature for the main 
grain species.
Note that the calculated dust temperature is not sensitive to grain 
radius for the size range ($a \la 0.01$ $\mu$m) of dust formed in SNe Ia.
The temperatures of MgSiO$_3$, Mg$_2$SiO$_4$, and SiO$_2$ grains have 
been lower than the gas temperature ($T / T_{d,j} > 1$) since their 
formation and are as low as $T_{d,j} =$ 90--150 K at 400 days.
On the other hand, Fe, FeS, and Si grains hold $T_{d,j} \sim$ 600 K, 
720 K, and 1000 K at 400 days, respectively, and have kept higher 
temperatures than the gas ($T / T_{d,j} < 1$).
The temperature of C grains becomes higher or lower than the gas 
temperature, depending on position and time.

Using the calculated time evolution of dust temperature and taking 
account of the effect of non-local thermal equilibrium, we performed 
the dust formation calculations for C, Fe, FeS, and Si grains with 
$T / T_{d,j} < 1$.
As the results of calculations, the condensation of Fe, FeS, and Si 
grains is delayed by 50--350 days, and their condensation times are 
300--400 day for Fe and FeS grains, and 450--500 days for Si grains.
In such late times, the gas density is too low for dust grains to 
form efficiently, so the resulting grain masses are heavily reduced: 
$\sim$$5 \times 10^{-6}$ $M_\odot$ for Fe grains, 
$\sim$$4 \times 10^{-4}$ $M_\odot$ for FeS grains, and 
$\sim$$1 \times 10^{-6}$ $M_\odot$ for Si grains 
in both the models A1 and B1.
Thus, apart from Fe grains formed in the innermost ejecta, FeS and Si 
grains make a negligible contribution to the opacity 
($\tau_{0.55, j} \le 0.1$) after $\ge$ 300 days.
On the other hand, the mass of C grains, whose condensation times are 
5--20 days later than those for the case of the local thermal equilibrium, 
decreases only by less than $\sim$10 \% for all of the models considered 
in this paper.
Namely, too high optical depth ($\tau_{0.55, C} \ge 10$) endowed by 
C grains cannot be mitigated by the effects of the non-local thermal 
equilibrium.

It should be mentioned here that, using a kinetic approach of 
nucleation, Lazzati (2008) pointed out that the nucleation of small 
clusters with a temperature lower than that of the gas can proceed 
faster in the absence of a strong radiation field. 
However, to examine the stability of small clusters in the SN ejecta 
with a strong radiation field, the nanoscale effect such as 
thermal fluctuations and the finite energy of small clusters need to 
be treated properly.
Such a computation is too time-consuming (Keith \& Lazzati 2011) and 
is beyond the scope of this paper.

\subsection{Thermal Emission from Newly Formed Dust}

In this subsection, we assess the thermal emission spectra from dust 
formed in the ejecta by employing the equilibrium temperature of dust 
given in the last subsection.
Then, we compare the results with the observed mid-IR SEDs and discuss
the mass and composition of newly formed dust.
The derived IR spectra will be also useful as a guide to the future 
IR observations of SNe Ia.

The flux densities of thermal emission from newly formed dust are 
calculated with Equations (2) and (3) in Nozawa et al.\ (2008), by 
taking account of the self-absorption of dust emission.
In this calculation, we adopt the mass of each dust species resulting 
from the dust formation calculations under the non-local thermal 
equilibrium condition for the models A1 and B1 (hereafter the models 
A1-non and B1-non, respectively).
The observational data are taken from the photometric results of SN 
at $\simeq$200 days and $\simeq$400 days by Gerardy et al.\ 
(2007), which may be the only mid-IR measurements of normal SNe Ia 
at late times.
Gerardy et al.\ (2007) claimed that the observed mid-IR spectra are 
not due to thermal emission from dust but are dominated due to line 
emissions.
Here, we regard the observed flux densities as the upper limits for 
thermal emission from dust.
We take $D = 18.7$ Mpc as a distance to SN 2005df (Hamuy 2003).  

Figures 6(a) and 6(b) present the calculated total mid-IR SEDs at day 
200 and day 400, respectively; the thin solid (dashed) lines are the 
thermal emission spectra in the case where the contribution of C grains
is included for the model A1-non (B1-non).
In both models and in both dates, the thermal emission from C grains 
formed in the inner part of the C--O layer, as well as from 
silicate grains in the inner ejecta, is mostly absorbed, and the 
total IR SEDs are dominated by that from C grains in the outermost 
part of the C--O layer.
(As a result, the difference of the IR SEDs between the model A1-non 
and B1-non is small, even if the mass of C grains for the model B1-non 
is about one order of magnitude larger than for the model A1-non.)
It should be noted that the calculated flux densities are a few orders 
of magnitude higher than the observed values;
if C grains had actually condensed in SNe Ia 2005df, the IR 
observations would have detect their continuum emission. 
This outcome, as well as the very high optical depths, conflicts
the observations, implying that the formation of C grains in SNe Ia 
should be extremely inefficient or unsuccessful.

%%%%%%%%%%%%%%%%%%%%%%%%%%%%%%%%%%
\begin{figure}
%\epsscale{0.6}
\epsscale{1.0}
\plotone{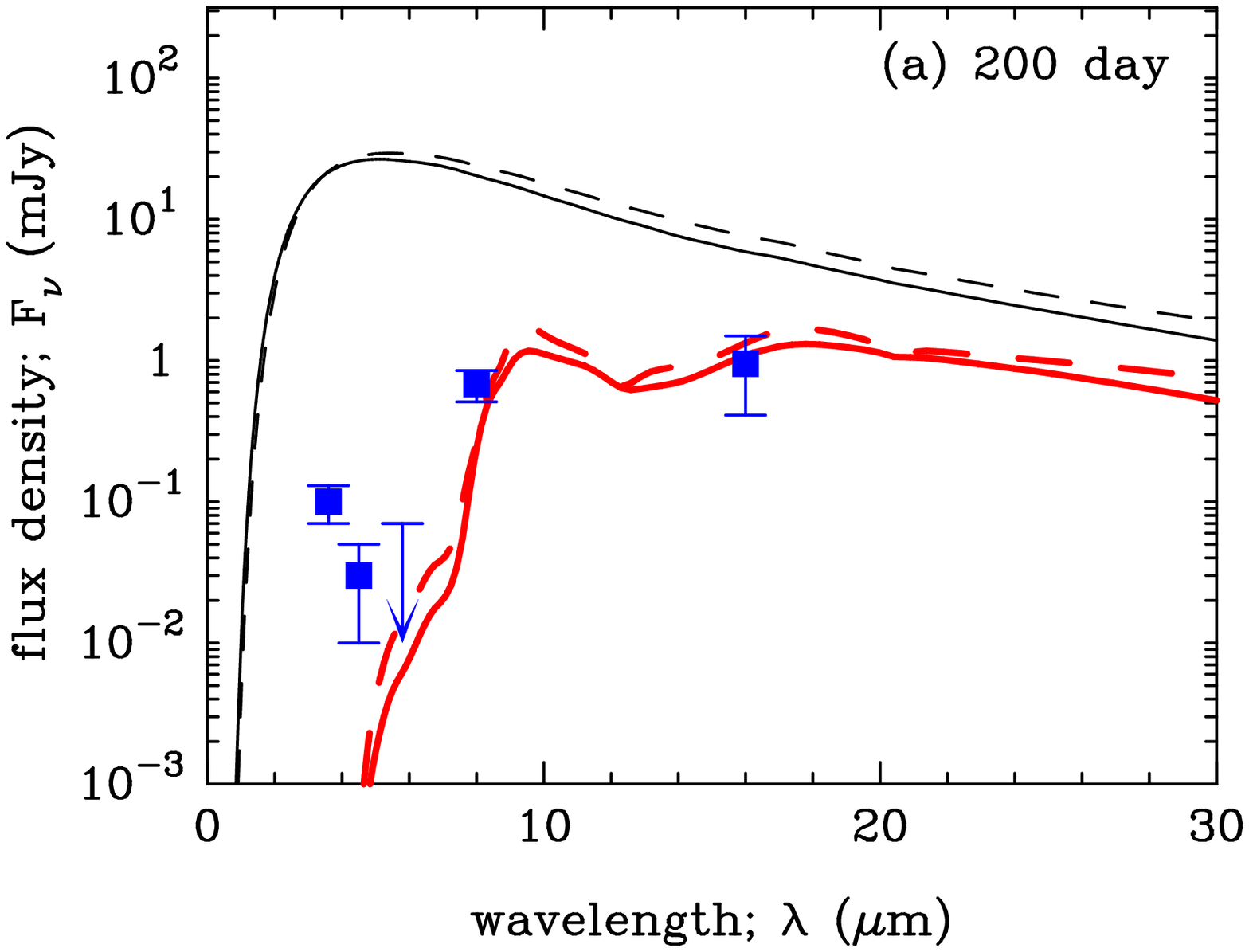}
\plotone{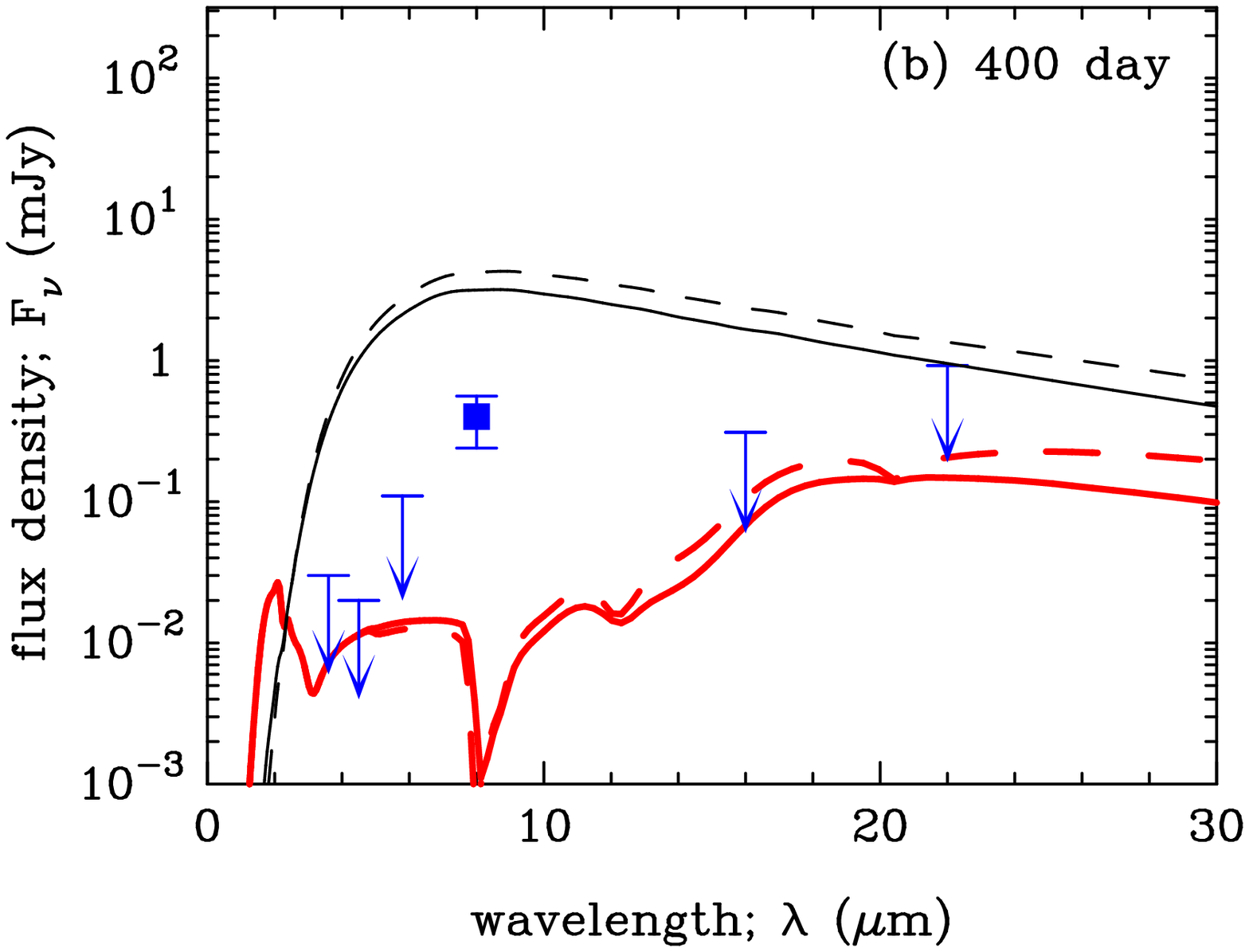}
\caption{
 Spectral energy distributions by thermal emission from newly formed 
 dust for the models A1-non (solid lines) and B1-non (dashed lines) at 
 (a) 200 days and (b) 400 days after the explosion.
 The thin lines are the thermal emission spectra summed up over all of 
 the grain species formed in the ejecta before 200 days or 400 days.
 The thick lines are the mid-IR spectra obtained by assuming the 
 absence of C grains.
 The observational data are the photometric results of SN 2005df
 by the {\it Spitzer} and are taken from Gerardy et al.\ (2007).
\label{fig6}}
\end{figure}
%%%%%%%%%%%%%%%%%%%%%%%%%%%%%%%%%%

In order to get further insight into the composition and mass of dust 
formed in SNe Ia, we also calculated the mid-IR spectra in the case 
where the contribution of C grains is removed arbitrarily.
The results are drawn by the thick lines in Figures 6(a) and 6(b).
In the models A1-non (solid lines) and B1-non (dashed lines), the 
IR SEDs at 200 day are produced by silicate grains with the mass of
0.03 $M_\odot$ and 0.075 $M_\odot$, respectively.
At 400 days, in addition to the contribution of silicate grains at 
$\lambda \ge 8$ $\mu$m, FeS grains of $\sim$4 $\times 10^{-4}$ 
$M_\odot$ contribute to the IR fluxes at $\lambda < 8$ $\mu$m.
As can be seen from these figures, the calculated SEDs do not 
contradict with the observational data at both day 200 and day 400.
Therefore, from this comparison, we cannot impose any constraints on
the composition and mass of grains species other than C grains.
In other words, $\sim$0.03--0.075 $M_\odot$ of silicate grains can 
be allowed as a mass of dust formed in SNe Ia within the present 
observational constraints.

\subsection{C grain formation and the Outermost C--O Layer of SNe Ia}

The discussion in the last subsection indicates that the formation of 
appreciable amounts of C grains in SNe Ia is incompatible with the 
existing observational results.
Thus, the condensation of C grains must be suppressed much more than 
the present calculations predict.
In our model, one of the factors that produces massive C grains is 
relatively high sticking probabilities of $\alpha_j \ge 0.1$;
if we take $\alpha_j \la 0.01$, neither C grains nor the other grains
can condense substantially.
Such low sticking probabilities are suggested for the formation of C 
grains in the peculiar Type Ib SN 2006jc;
Nozawa et al.\ (2008) showed that $\alpha_j \la 3 \times 10^{-3}$ 
is required for C grains to obtain the dust mass needed to reproduce 
the IR observations of SN 2006jc.
They argued that in SN 2006jc C grains can form at very early times of 
$\sim$50 days, when energetic photons and electrons are expected to 
be abundant, so the low sticking probability may reflect the effective 
destruction of small clusters by collisions with photons and electrons.
Given that the formation of C grains in SNe Ia can also occur at an 
early epoch of $\sim$100 days, it can be possible for precursor 
clusters of C grains to be destroyed significantly.

Another reason for the formation of a large quantity of C grains is 
the presence of massive ($\simeq$0.05 $M_\odot$) unburned carbon in 
the outer layer.
In the W7 model adopted in this paper, the deflagration wave halts on 
the way, and the original composition of the WDs with a number ratio 
of C/O $= 1.27$ is retained in the outermost ejecta.
However, whether the C/O ratio exceeds unity is not conclusive, and 
the composition of the outer layer of accreting WDs is unclear because 
of many uncertain processes such as the $^{12}$C$(\alpha, \gamma)^{16}$O 
reaction rate, convective overshooting, dredge-up from the core, 
accretion rate from the companion star, and strength of He-shell flashes
(e.g., Shen \& Bildsten 2009; Y.\ Kamiya et al.\ in preparation);
if the C/O number ratio is less than unity in the surface layer of the 
WD and if a majority of C atoms should be locked in CO molecules, C 
grains may be unable to form efficiently in the cooling ejecta.
Thus, observations of CO molecules in SNe Ia can provide meaningful 
clues to the formation process of C grains as well as the pre-explosion 
composition of the WDs, although the existence of CO molecules has 
never been reported for normal SNe Ia to date.

On the other hand, the mass of carbon deduced from the observations
is much smaller than in the W7 model.
From the analyses of the spectra before the maximum of the light 
curves (Marion et al.\ 2006; Tanaka et al.\ 2008), the mass of 
carbon in normal SNe Ia is estimated to be at most 0.01 $M_\odot$.
If, as mentioned in these studies, the abundance of carbon is 
10--100 times less than oxygen, the mass of newly formed C grains may 
be too low for observational signatures of dust formation to be detected. 
Therefore, from a point of view of dust formation, we assert that the 
composition of the outermost ejecta is very carbon-poor.
Such a small carbon abundance might result from strong He-shell flashes 
in accreting high-mass WDs (Y.\ Kamiya et al.\ in preparation), or be 
the results of the delayed detonation which burns almost all carbon.

Apart from the outermost layer, the one-dimensional delayed detonation 
models predict the temperature and density structures and the elemental 
composition well similar to those in the one-dimensional deflagration 
models (Iwamoto et al.\ 1999).
This implies that in the delayed detonation models the average radii 
and masses of newly formed grains except for C grains cannot 
significantly be different from those obtained in this paper.
The remaining amount of unburned carbon in the delayed detonation models
is smaller ($\sim$0.005--0.03 $M_\odot$) than 0.0475 $M_\odot$ 
in the deflagration model W7, but it heavily depends on the still 
uncertain transition density at which the deflagration turns into 
the detonation.
Thus, in order to investigate how the mass of newly formed C grains can 
be changed by the underlying explosion models, it will be necessary that 
comprehensive calculations of dust formation are performed for the 
delayed detonation models with different transition densities.

Here it would be worth mentioning dust formation in the extremely 
luminous SNe Ia.
Recently, four over-luminous SNe Ia have been discovered and suggested 
to originate from the progenitors with super-Chandrasekhar masses:
SN 2003fg (Howell et al.\ 2006), 
SN 2006gz (Hicken et al.\ 2007),
SN 2007if (Scalzo et al.\ 2010; Yuan et al.\ 2010), and
SN 2009dc (Yamanaka et al.\ 2009; Tanaka et al.\ 2010; Silverman et 
al.\ 2011).
Interestingly, these bright SNe Ia show much stronger C lines in their 
spectra near the maximum brightness than any other normal SNe Ia.
This implies that thick unburned C-rich layers are left in their 
outermost ejecta.
Since the formation of C grains accompanies the presence of the C 
layer as shown in this paper, C grain may be able to condense in these 
super-Chandra SNe Ia.
Indeed, Maeda et al.\ (2009) reported that SN 2006gz after one year is 
quite fainter than expected from the peak luminosity, suggesting 
that such a fading might be caused by dust formation.
Taubenberger et al.\ (2010) also discovered an enhanced fading of the 
light curve around 200 days for SN 2009dc.

The super-Chandra SNe Ia have high ejecta masses and the low 
expansion velocities.
Thus, the density of the ejecta is higher than those in normal SNe Ia, 
which may make dust formation more feasible.
On the other hand, energy deposition from a large amount of $^{56}$Ni 
($>$1 $M_\odot$) produced in super-Chandra SNe Ia prevents the gas 
temperature from quickly decreasing down to the dust condensation 
temperature ($\la$2000 K), which may make the condensation of dust 
unfeasible.
In order to clarify how these competing processes affect the 
condensation time and mass of dust, the formation of dust in 
super-Chandra SNe Ia should be explored both theoretically and 
observationally.

\section{Injection of Newly Formed Dust into the ISM}

The composition, size, and amount of dust grains injected from SNe 
into the ISM are determined by the conflicting processes between 
the formation of dust in the expanding SN ejecta and the subsequent 
destruction of the dust in the hot gas swept up by the reverse and 
forward shocks propagating within the SNRs (Nozawa et al.\ 2007, 2010; 
Bianchi \& Schneider 2007; Nath et al.\ 2008; Silvia et al.\ 2010).
In this section, in order to reveal the mass of dust ejected from SNe 
Ia, we investigate the evolution of dust in the shocked gas inside 
SNRs, applying the results of dust formation calculations.

The calculations of dust evolution in SNRs are based on the model by 
Nozawa et al.\ (2007, 2010).
In the models, the transport and destruction of dust within SNRs are
pursued simultaneously with the time evolution of the gas temperature 
and density in the spherical symmetry shocks;
by treating dust grains as test particles, the deceleration by the gas 
drag and the erosion by sputtering are self-consistently calculated,
according to the initial radius and the initial position of each dust 
species.
The sputtering yield of each dust species is taken from Nozawa et 
al.\ (2006).

Note that the destruction efficiency of dust heavily depends on the 
ambient gas density as well as the initial grain radius.
In the present calculations, we assume that the ISM around SNe Ia is 
uniform, and consider the hydrogen number density of $n_{{\rm H},0} =$ 
0.01, 0.1, and 1 cm$^{-3}$ as the gas density in the ISM, whose gas 
composition is set as the solar abundance.
As for the model of dust formed in the ejecta of SNe Ia, we adopt the 
results for the models A1 and B1 given in Section 3;
the adoption of these models is intended to estimate the possible 
maximum mass of each dust species supplied into the ISM.
Assuming that the ejecta interact with the ambient medium at 30 yr 
after the explosion, we follow the calculations up to a few $\times$
$10^6$ yr.

Figure 7(a) shows the time evolutions of the total mass of dust in 
SNRs for the models A1 (solid lines) and B1 (dashed line).
We can see that for $n_{{\rm H}, 0} \ge 0.1$ cm$^{-3}$ the dust grains 
formed in the ejecta of SNe Ia are completely ($\le$$10^{-5}$ $M_\odot$) 
destroyed before 10$^6$ yr.
Such effective destruction of dust arises from the following two
reasons.
(1) Since SNe Ia do not have H-envelope, the reverse shock can sweep up
the dust formation region much earlier ($<$500 yr) than in 
envelope-retaining SNe II-P ($>$1000 yr).
At such early times, the gas density in the shocked ejecta is high 
enough that dust grains are efficiently decelerated and eroded due 
to frequent collisions with the gaseous ions.
(2) Since the radii of newly formed grains are small ($\la$0.01 $\mu$m), 
they are quickly decelerated by the gas drag soon after encountering 
the reverse shock.
Thus, these small grains are finally trapped in the postshock gas and 
are destroyed by thermal sputtering without being injected into the ISM.

On the other hand, for the ISM density as low as $n_{{\rm H}, 0} = 0.01$ 
cm$^{-3}$, the deceleration and destruction of dust in the hot plasma
are quite inefficient, and dust of $\simeq$(1-2)$\times 10^{-2}$ 
$M_\odot$ survive in the models of A1 and B1;
C, FeS, and Mg$_2$SiO$_4$ grains with smaller average radii ($a_{\rm ave} 
\la 30$~\AA) are trapped in the shocked gas to be completely destroyed, 
whereas a fraction of Si, SiO$_2$, MgSiO$_3$, and Fe grains with larger
average radii ($a_{\rm ave} \ga 30$~\AA) are injected into the ISM.
However, the total mass of surviving dust is dominated by Si grains
(see Fig.\ 7(b) for the model A1), whose formation has been expected 
to be greatly suppressed by the effect of the non-local thermal 
equilibrium (see Section 4.2).
Thus, neglecting the contribution of Si grains, we regard 
$\sim$$10^{-3}$ $M_\odot$ ($\sim$$5 \times 10^{-3}$ $M_\odot$) of 
silicate grains as a more realistic mass estimate of surviving dust 
for $n_{{\rm H}, 0} = 0.01$ cm$^{-3}$ in the model A1 (B1).\footnote
{In the model B1, in addition to silicate grains of 
$\sim$$5 \times 10^{-3}$ $M_\odot$, C grains of $4 \times 10^{-3}$ 
$M_\odot$ can survive for $n_{{\rm H}, 0} = 0.01$ cm$^{-3}$.
However, as discussed in Section 4, the formation of C grains is 
problematic, judging from the comparison with the observational 
constraints.
Therefore, we do not include here the mass of C grains as an estimate 
of the total mass of surviving dust.}
These results indicate that, even if the ambient gas density is very 
low, the amount of dust ejected from SNe Ia into the ISM is 
considerably small.
We also note that typical ISM densities around SNe Ia are estimated to 
be $n_{{\rm H}, 0} =$ 1--5 cm$^{-3}$ (e.g., Borkowski et al.\ 2006) and
that the ambient density lower than $n_{{\rm H}, 0} \sim 0.01$ cm$^{-3}$ 
results in too extended SNRs whose sizes are incompatible with observed 
ones of nearby Type Ia SNRs (Badenes et al.\ 2007).

%%%%%%%%%%%%%%%%%%%%%%%%%%%%%%%%%%
\begin{figure}
%\epsscale{0.6}
\epsscale{1.1}
\plotone{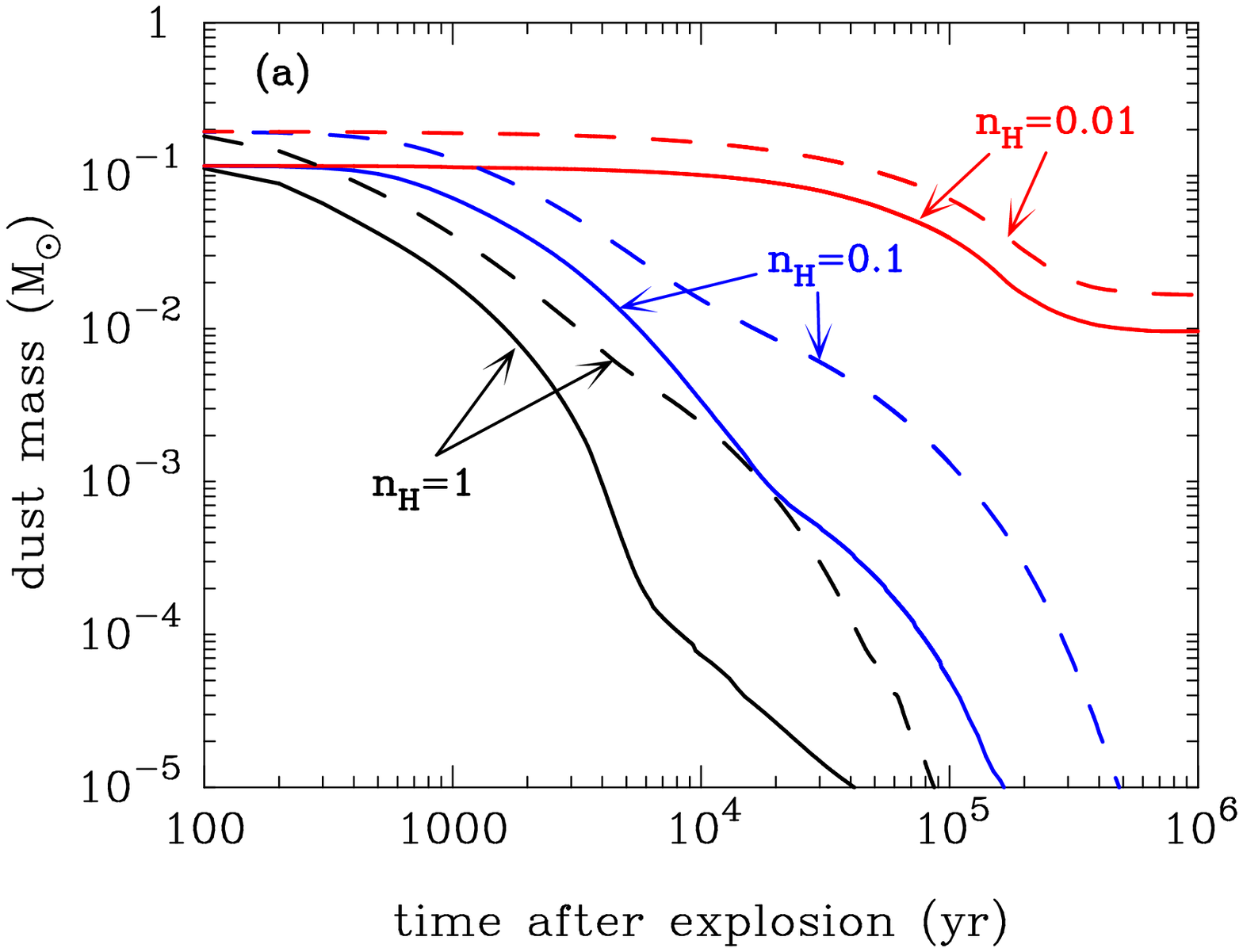}
\plotone{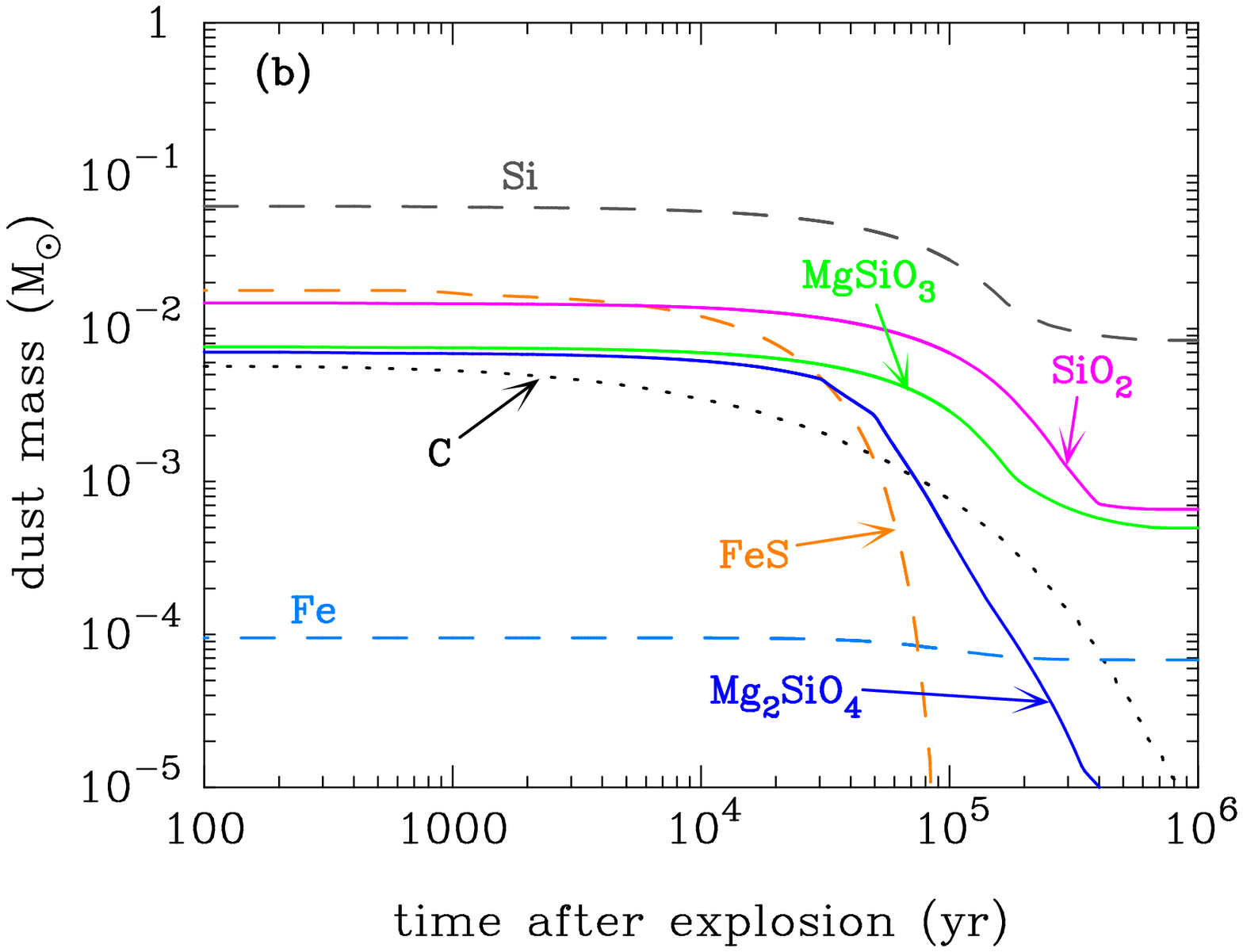}
\caption{
 (a) Time evolutions of the total mass of the newly formed dust within 
 the Type Ia SNRs expanding into the uniform ISM with the hydrogen number 
 density of $n_{{\rm H},0} =$ 0.01, 0.1, and 1 cm$^{-3}$.
 The solid lines are for the model A1, and the dashed lines are for 
 model B1.
 (b) Time evolutions of the mass of each dust species for the model A1 
 and for $n_{{\rm H},0} =$ 0.01 cm$^{-3}$.
\label{fig7}}
\end{figure}
%%%%%%%%%%%%%%%%%%%%%%%%%%%%%%%%%%

It should be mentioned that, as discussed in Nozawa et al.\ (2010), if 
SNRs expand non-spherically, a part of the newly formed grains may be 
able to evade the destruction and to be injected in the directions that 
do not strongly interact with the ambient medium.
Recently, a variation of late-time nebular spectra of SNe Ia has been
successfully explained by taking into account the global asymmetry in 
the innermost ejecta that is produced by deflagration developing from 
the off-center ignition (Maeda et al.\ 2010a, 2010b, see also 
Motohara et al.\ 2006).
Aside from the global asymmetry issue, the deflagration flame is also 
expected to create a small-scale mixing structure (e.g., R\"{o}pke et 
al.\ 2007).
In addition, given that SNe Ia are the end-products of the binary 
interaction, aspherical circumstellar medium might have been formed 
around the exploding WDs, which may result in the non-spherical 
evolution of SNRs.
However, the analyses of X-ray morphology of young SNRs (Lopez et al.\ 
2009, 2010) has showed that the remnants of SNe Ia do not seem to 
be far from spherical symmetry.
Although the effect of degree of non-sphericity and small-scale 
structure on dust survival is uncertain, we conceive here that these
effects are not enough to make SNe Ia dominant sources of interstellar 
dust.

\section{Summary}

We investigate the formation of dust in the ejecta of SNe Ia, adopting 
the carbon-deflagration W7 model.
In our calculations, we apply the nucleation and grain growth theory,
to compare with the results for dust formation in different types of 
CCSNe in our earlier studies.
We find that for the sticking probability of $\alpha_j =$ 1, various 
grain species can condense in the stratified ejecta of SNe Ia, although
Fe grains cannot form appreciably, contrary to the expectation.
The composition of dust grains formed in SNe Ia reflects the elemental
composition of the gas in the ejecta and is basically the same as 
those in any other CCSNe.

On the other hand, the condensation times of dust in SNe Ia are much 
earlier ($t_c =$ 100--300 days), and the average radius of newly formed 
dust are much smaller ($a_{\rm ave} \la 0.01$ $\mu$m) than those in SNe
II-P.
This is due to the low gas density in the ejecta of SNe Ia that do not
have H-rich envelopes.
We conclude that the radius of newly formed dust depends on the type 
of SNe and that smaller grains condense in SNe with less massive 
envelopes. 
The total mass of dust that can condense in the ejecta of SNe Ia is 
significantly affected by the sticking probability $\alpha_j$ and the 
formation efficiency of CO and SiO molecules, ranging from 
$3 \times 10^{-4}$ $M_{\odot}$ to 0.2 $M_{\odot}$ for $\alpha_j =$ 
0.1--1.

Furthermore, we estimate the temperature of small clusters and 
evaluate the effect of the non-local thermal equilibrium on the 
formation process of dust.
We find that the non-local thermal equilibrium effect can suppress 
the condensation of FeS, Si, and Fe grains that are otherwise the main 
donors to the optical depths.
We also calculate the IR emission spectra from the newly formed dust,
which would be helpful in discussing the possibility of dust formation 
from the future observations of SNe Ia.
On the other hand, from the present observational constraints, we 
notice that the formation of massive C grains is suppressed in SNe Ia. 
This implies that the nucleation of C grains fails due to the 
destruction by energetic photons and electrons, or that the outermost 
C-O layer of SNe Ia are burned by the delayed detonation wave, or
that the pre-supernova C/O ratio is exceedingly small as a result of
strong recurrent He-shell flashes.

Finally, we examine the survival of dust grains formed in the ejecta 
against their destruction in the SNRs.
We find that, unless the gas density around SNe Ia is too low 
($n_{{\rm H}, 0} \le 0.01$ cm$^{-3}$), the newly formed grains are 
almost completely destroyed in the shocked gas in the course of their 
injection into the ISM.
Even if the asymmetric effect of explosion is taken into account, 
SNe Ia is likely to be poor producers of interstellar dust.
However, SNe Ia are major production factories of heavy elements which
will be available for the subsequent growth of the pre-existing grains 
through accretion onto their surface in the dense molecular clouds.

\acknowledgments

We thank R. Kotak and Y. Kamiya for the useful comments.
We are grateful to the anonymous referee for critical comments
that are useful for improving the manuscript.
This research has been supported in part by World Premier International 
Research Center Initiative (WPI Initiative), MEXT, Japan, and by the 
Grant-in-Aid for Scientific Research of the Japan Society for the 
Promotion of Science (18104003, 20340038, 22684004, 22840009).

%\newpage

\end{document}